\documentclass{pazhbeng}

\usepackage{graphicx}
\usepackage{floatrow}
\usepackage{longtable}

\usepackage[hyperfootnotes=false,draft]{hyperref}
 
\def\smfigure#1#2#3{
 \begin{minipage}{0.47\columnwidth}
    \begin{minipage}{0.049\columnwidth}
      \rotatebox{90}{\footnotesize\phantom{0000}#3}
    \end{minipage}
    \begin{minipage}{0.9\columnwidth}
     \includegraphics[viewport=50 188 700 500,clip,width=1.5\columnwidth]{#1}
      \centerline{\footnotesize #2}
    \end{minipage}

    ~
  \end{minipage}
}

\def\smfiguresmall#1#2#3{
  \begin{minipage}{0.47\textwidth}
    \begin{minipage}{0.049\columnwidth}
      \rotatebox{90}{\footnotesize\phantom{0000}#3}
    \end{minipage}
    \begin{minipage}{0.9\columnwidth}
     \includegraphics[viewport=50 188 700 500,width=1.30\columnwidth]{#1}
      \centerline{\footnotesize #2}
    \end{minipage}
  \end{minipage}
}

\def\doubleline{\vskip 3pt\hrule \vskip 1.5pt \hrule \vskip 5pt}

\def\ps1{\emph{Pan-STARRS1}}

\def\srg{\textit{SRG}}
\def\art{ART-XC}
\def\ero{eROSITA}
\def\rosat{\textit{ROSAT}}
\def\xmm{\textit{XMM-Newton}}
\def\swift{\textit{Swift}}

\def\nh{N_{\rm H}}
\def\lx{L_{\rm X}}
\def\mbh{M_{\rm BH}}
\def\ledd{L_{\rm Edd}}
\def\lbol{L_{\rm bol}}

\begin{document}

%\journalinfo{2020}{0}{0}{1}[0]
\journalinfo{2021}{47}{2}{71}[87]
\UDK{524.77}

\title{OPTICAL IDENTIFICATION OF ACTIVE GALACTIC NUCLEUS CANDIDATES DETECTED BY THE MIKHAIL PAVLINSKY \art\ TELESCOPE ABOARD THE \srg\ OBSERVATORY DURING THE ALL-SKY X-RAY SURVEY}

\author{I.A.~Zaznobin\address{1}\email{zaznobin@iki.rssi.ru},
  G.S.~Uskov\address{1},
  S.Yu.~Sazonov\address{1},
  R.A.~Burenin\address{1},
  P.S.~Medvedev\address{1},
  G.A.~Khorunzhev\address{1},
  A.R.~Lyapin\address{1},
  R.A.~Krivonos\address{1},
  E.V.~Filippova\address{1},
  M.R.~Gilfanov\address{1,2},
  R.A.~Sunyaev\address{1,2},
  M.V.~Eselevich\address{3},
  I.F.~Bikmaev\address{4,5},
  E.N.~Irtuganov\address{4},
  E.A.~Nikolaeva\address{4}
  \addresstext{1}{Space Research Institute, Profsoyuznaya str. 84/32, Moscow, 117997 Russia}
  \addresstext{2}{Max Planck Institut fur Astrophysik, Karl-Schwarzschild-Str. 1, Postfach 1317, D-85741 Garching, Germany}
  \addresstext{3}{Institute of Solar–Terrestrial Physics, Russian Academy of Sciences, Siberian Branch, P.O. Box 291, Irkutsk, 664033 Russia}
  \addresstext{4}{Kazan Federal University, Kremlevskaya str. 18, Kazan, 420000 Russia}
  \addresstext{5}{Academy of Sciences of Tatarstan, Baumana str. 20, Kazan, Russia}
}

\shortauthor{}

\shorttitle{}

\submitted{November 26, 2020}

\begin{abstract}

We present the results of our identification of eight objects from a preliminary catalogue of X-ray sources detected in the 4--12 keV energy band by the Mikhail Pavlinsky \art\ telescope aboard the \srg\ observatory during its first all-sky survey. Three of them (SRGA\,J$005751.0\!+\!210846$, SRGA\,J$014157.0\!-\!032915$, SRGA\,J$232446.8\!+\!440756$) have been discovered by \art, while five were already known previously as X-ray sources, but their nature remained unknown. The last five sources have also been detected in soft X-rays by the \ero\ telescope of the \srg\ observatory. Our optical observations were carried out at the 1.6-m AZT-33IK telescope of the Sayan Observatory and the 1.5-m Russian–Turkish telescope (RTT-150). All of the investigated objects have turned out to be active galactic nuclei (AGNs) at redshifts from 0.019 to 0.283. Six of them are Seyfert 2 galaxies (including one Seyfert 1.9 galaxy), one (SRGA\,J$005751.0\!+\!210846$) is a “hidden” AGN (in an edge-on galaxy), and one (SRGA\,J$224125.9\!+\!760343$) is a narrow-line Seyfert 1 galaxy. The latter object is characterized by a high X-ray luminosity ($\sim (2-13)\times 10^{44}$\,erg\,s$^{-1}$ in the 4--12 keV band) and, according to our black hole mass estimate ($\sim 2\times 10^7 M_\odot$), an accretion rate close to the Eddington limit. All three AGNs discovered by the \art\ telescope (which are not detected by the \ero\ telescope) are characterized by a high absorption column density ($\nh\gtrsim 3\times 10^{23}$\,cm$^{-2}$). The results obtained confirm the expectations that the \art\ telescope is an efficient instrument for searches of heavily obscured and other interesting AGNs in the nearby ($z\lesssim 0.3$) Universe. The SRG all-sky survey will last for more than 3 years more, which will allow many such objects to be discovered.


\end{abstract}

\section{INTRODUCTION}
The Russian Mikhail Pavlinsky \art\ telescope \citep{artxc} aboard the Russian \srg\ orbital observatory  \citep{srg21} has been conducting an all-sky X-ray survey at energies from 4 to 30 keV since December 2019. Mirrors operating on the principle of grazing X-ray incidence and semiconductor detectors based on cadmium telluride crystals are used in the telescope, providing unique characteristics for this energy range: a wide field of view (36 arcmin) and good angular resolution (better than 1 arcmin in the sky scanning mode). Because of this, during the four-year survey we expect to obtain an all-sky map unique in depth and sharpness at energies 4--12 keV and, in particular, to detect at least 5000 active galactic nuclei (AGNs), which is several times more than has been found at such energies in previous all-sky surveys.

In June 2020, the \srg\ observatory completed the first (of the planned eight) all-sky scans, and a preliminary catalogue of detected sources (more than 600 objects in total) was produced from the \art\ data obtained. This catalogue was correlated with: (1) catalogues of sources detected in previous X-ray sky surveys; (2) the preliminary catalogue of sources detected on one half of the celestial sphere $0<|l|<180^\circ$\footnote{Russian scientists are responsible for processing the data from the \ero\ telescope (Germany) in this part of the sky.} in the soft X-ray energy band during the first \srg/\ero\ survey; (3) catalogues of astrophysical objects in other wavelength ranges (from radio to ultraviolet). As a result, a list of objects consisting of the sources discovered by the \art\ telescope and previously known X-ray sources of unknown nature was compiled. Some of these objects were also detected by the \ero\ telescope \citep{pred20} of the \srg\ observatory.

Spectroscopic observations are carried out at Russian optical telescopes to identify these potentially interesting \art\ sources. The first results of this observational campaign are presented in the present paper. The eight \art\ sources to be discussed below have proved to be type 1 or 2 AGNs, including objects with strong internal absorption. The latter was revealed by analyzing the X-ray spectra constructed from the \art\ and \ero\ data.

The presented luminosity estimates are based on the model of a flat Universe with parameters $H_0=70$ and $\Omega_m = 0.3$.

\section{THE SAMPLE OF OBJECTS}

The objects being studied (see Table~\ref{tab:list}) were selected among the point X-ray sources detected by the \art\ telescope during the first sky survey (December 12, 2019–June 10, 2020) with a signal-to-noise ratio of no less than 4.5 in the 4--12 keV energy band. Based on the \art\ data, we measured the positions of the sources in the sky and their fluxes in this energy band. The typical position error (at 95\% confidence) is 30 arcsec.

For these objects, based on the \ero\ data, we obtained the fluxes or upper limits on the flux in three energy bands: 0.3–2, 2–6, and 4–9 keV. Out of the eight sources detected by the \art\ telescope, 5 were also detected by the \ero\ telescope either in all three or in the first two of these bands. For them, based on the \ero\ data, we managed to improve the source position in the sky. The remaining three objects are not detected by the \ero\ telescope.

For all our objects Table~\ref{tab:list} gives: the coordinates of the \art\ source, the coordinates of the putative optical counterpart, the distance between the position of the optical counterpart and the positions of the X-ray source from the \art\ and \ero\ data (if available), and the X-ray observatory that first discovered the X-ray source.

\begin{table*}
  \caption{List of objects for our spectroscopic observations}
  
  \label{tab:list}
  \vskip 2mm
  \renewcommand{\arraystretch}{1.1}
  \renewcommand{\tabcolsep}{0.35cm}
  \centering
  \footnotesize
  \begin{tabular}{lcclll}
    \noalign{\doubleline}     
    & \multispan2\hfil Optical coordinates\hfil & \\   
    \art\ source & $\alpha$ & $\delta$ & $r$ (\art) & $r$ (\ero) & Discovered by \\
    \noalign{\vskip 3pt\hrule\vskip 5pt}
    SRGA\,J$005751.0\!+\!210846$ & $00~57~52.1$ & $+21~08~46$ & $15.4\arcsec$ & -- & \srg\ \\
    SRGA\,J$014157.0\!-\!032915$ & $01~41~59.4$ & $-03~29~34$ & $40.6\arcsec$ & -- & \srg\ \\
    SRGA\,J$043209.6\!+\!354917$ & $04~32~08.0$ & $+35~49~29$ & $22.9\arcsec$ & $2.3\arcsec$ & \rosat\ \\
    SRGA\,J$045049.8\!+\!301449$ & $04~50~48.0$ & $+30~15~03$ & $27.2\arcsec$ & $3.2\arcsec$ & \swift\ \\
    SRGA\,J$152102.3\!+\!320418$ & $15~21~01.8$ & $+32~04~14$ & $7.5\arcsec$ & $2.9\arcsec$ & \swift\ \\
    SRGA\,J$200431.6\!+\!610211$ & $20~04~32.4$ & $+61~02~31$ & $20.8\arcsec$ & $5.3\arcsec$ & \rosat\ \\
    SRGA\,J$224125.9\!+\!760343$ & $22~41~25.8$ & $+76~03~53$ & $10.0\arcsec$ & $4.6\arcsec$ & \rosat\ \\
    SRGA\,J$232446.8\!+\!440756$ & $23~24~48.4$ & $+44~07~57$ & $17.3\arcsec$ & -- & \srg\ \\
    \noalign{\vskip 3pt\hrule\vskip 5pt}
  \end{tabular}

  \begin{flushleft}
  Column 1: the source name in the preliminary \art\ catalogue (the coordinates of the X-ray sources used in the names are given for epoch J2000.0). Columns 2 and 3: the coordinates of the putative optical counterpart. Column 4: the distance between the \art\ source position and the optical counterpart position. Column 5: the separation between the \ero\ source position and the optical counterpart position (the dash means that a given source is not detected by the \ero\ telescope). Column 6: the X-ray observatory that discovered the source.
  \end{flushleft}
\end{table*}

\section{X-RAY OBSERVATIONS}

The X-ray radiation from AGNs can experience photoabsorption in the gas–dust torus around the supermassive black hole (SMBH) and in the interstellar medium of the host galaxy. One of the goals of this study was to estimate the column density of neutral (or weakly ionized) matter $\nh$ for the objects being discussed. Although the number of X-ray photons detected by the \art\ and \ero\ telescopes (in the short source scanning time during the SRG sky survey) is insufficient to perform a detailed spectral analysis, these data nevertheless allow sufficiently reliable constraints on $\nh$ to be obtained in most cases.

We fitted the X-ray spectra in the range 0.3–12 keV using the XSPEC v12.9.0n2\footnote{https://heasarc.gsfc.nasa.gov/xanadu/xspec/} code jointly based on the \art\ and \ero\ data. The spectra were first binned in such a way that there were at least three counts in each spectral bin.

We assumed the AGN X-ray spectrum to be fitted by a power law with a fixed slope $\Gamma=1.8$ (a typical value for Seyfert galaxies) and a low-energy cutoff as a result of photoabsorption in the Galaxy and the object itself. Thus, we used the following model in XSPEC:
\[
phabs(zphabs(powerlaw)),
\]
where phabs is the absorption in the Galaxy from HI4PI survey data \citep{bekhti16} and zphabs is the absorption at the AGN redshift z (measured from the object’s optical spectrum). A satisfactory quality of the fit was achieved for all of the sources.

The X-ray spectra obtained are presented on the graphs below in units of $F_E(E)$. A power-law model with a slope $\Gamma=1.8$ was used to convert the counts on the detector to photons. It should be kept in mind that such figures should not be used to obtain accurate fluxes.

\section{Optical observations}

Our spectroscopy for the objects was performed at the 1.6-m AZT-33IK telescope of the Sayan Observatory using the low- and medium-resolution ADAM spectrograph \citep{adam16,adam16a} and the 1.5-m Russian–Turkish telescope (RTT-150) using the \emph{TFOSC}\footnote{http://hea.iki.rssi.ru/rtt150/en/index.php?page=tfosc} spectrograph. A set of long slits is used at both spectrographs to obtain the spectra.

We used volume phase holographic gratings (VPHG), 600 lines per millimeter, to obtain the spectra at the ADAM spectrograph. As a dispersive element we used VPHG600G for the spectral range 3650--7250~\AA\ with a resolution of 4.3~\AA\ and VPHG600R for the spectral range 6460--10\,050\,\AA\ with a resolution of 6.1~\AA. When using VPHG600R, we set the OS11 filter, which removes the second interference order from the image. A thick e2v CCD30-11 array produced by the deep depletion technology is installed at the spectrograph. This allows the spectral images to be obtained at a wavelength of 1 $\mu m$ without interference on the thin CCD substrate. A set of slits is available at the spectrograph; we used a 2\arcsec -wide slit to obtain the spectroscopic images. All our observations were performed with zero slit position angle. After each series of spectroscopic images for each object, we obtained the calibration images of a lamp with a continuum spectrum and the line spectrum of a He–Ne–Ar lamp.

Transmitting diffraction grating no. 15 with the spectral range 3700--8700\,\AA, which provides a spectral resolution of 12\,\AA, was used at the TFOSC spectrograph as a dispersive element. This grating allows bright Balmer lines to be obtained in the spectral images for galaxies up to $z=0.32$. To obtain the spectroscopic images, we used a 2\arcsec -wide slit. The spectrograph slit position angle is $90^{\circ}$. Before and after obtaining the series of spectroscopic images for each object, we obtained the images of a lamp with a continuum spectrum and the line spectrum of a Fe–Ar lamp.

All our observations were carried out at dark moonless time. Before obtaining the spectroscopic images, we tried to place the galactic nucleus at the center of the spectrograph slit as accurately as possible. After each exposure, we changed the object’s position along the slit by 10–15\arcsec\ in a random direction upward or downward using a photoguide. On each night at both telescopes we took the spectra of spectrophotometric standards from the ESO\footnote{https://www.eso.org/sci/observing/tools/standards/spec- tra/stanlis.html} list for all the sets of diffraction gratings and slits being used. The data reduction was performed using IRAF\footnote{http://iraf.noao.edu/} software package and our own software.

To estimate the broadening of emission lines, their profiles were fitted by a Gaussian, with the back-
ground having been fitted by a polynomial. The line width was defined as ${FWHM} = \sqrt{{FWHM}^2_{\rm mes} - {FWHM}^2_{\rm res}}$, where ${FWHM}_{\rm mes}$ is the measured line width and ${FWHM}_{\rm res}$ is the spectral broadening of the instrument whose values were given above for each dispersive element being used.

We classified the Seyfert galaxies based on their optical spectra in a standard way \citep{oster, veron}).

\begin{figure*}
  \centering
  \vfill
  SRGA\,J$005751.0\!+\!210846$
  \vfill
  \vskip 0.5cm
  \begin{floatrow}
    \includegraphics[width=0.4\columnwidth]{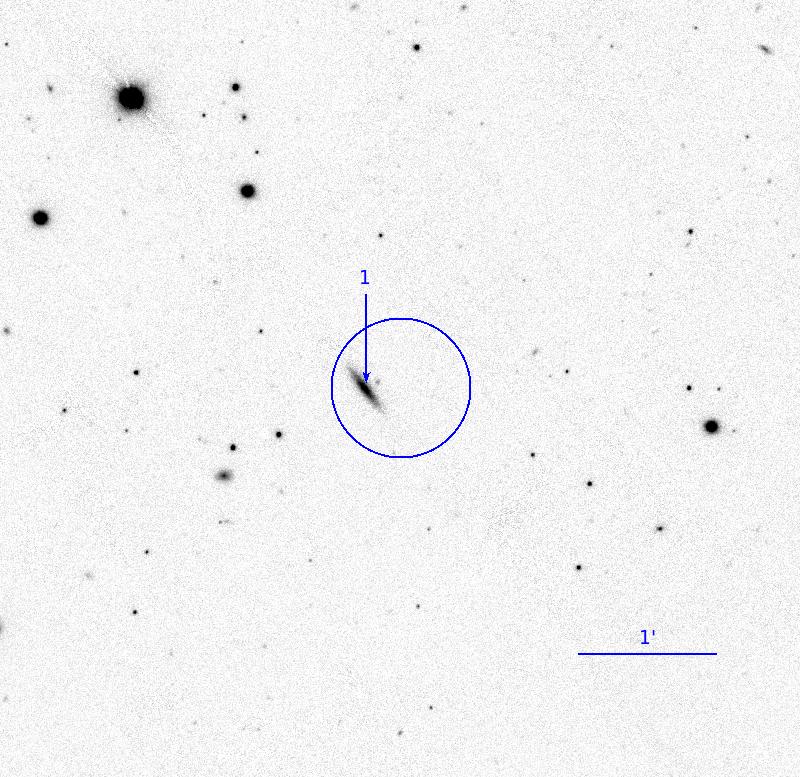}
    \includegraphics[width=0.4\columnwidth]{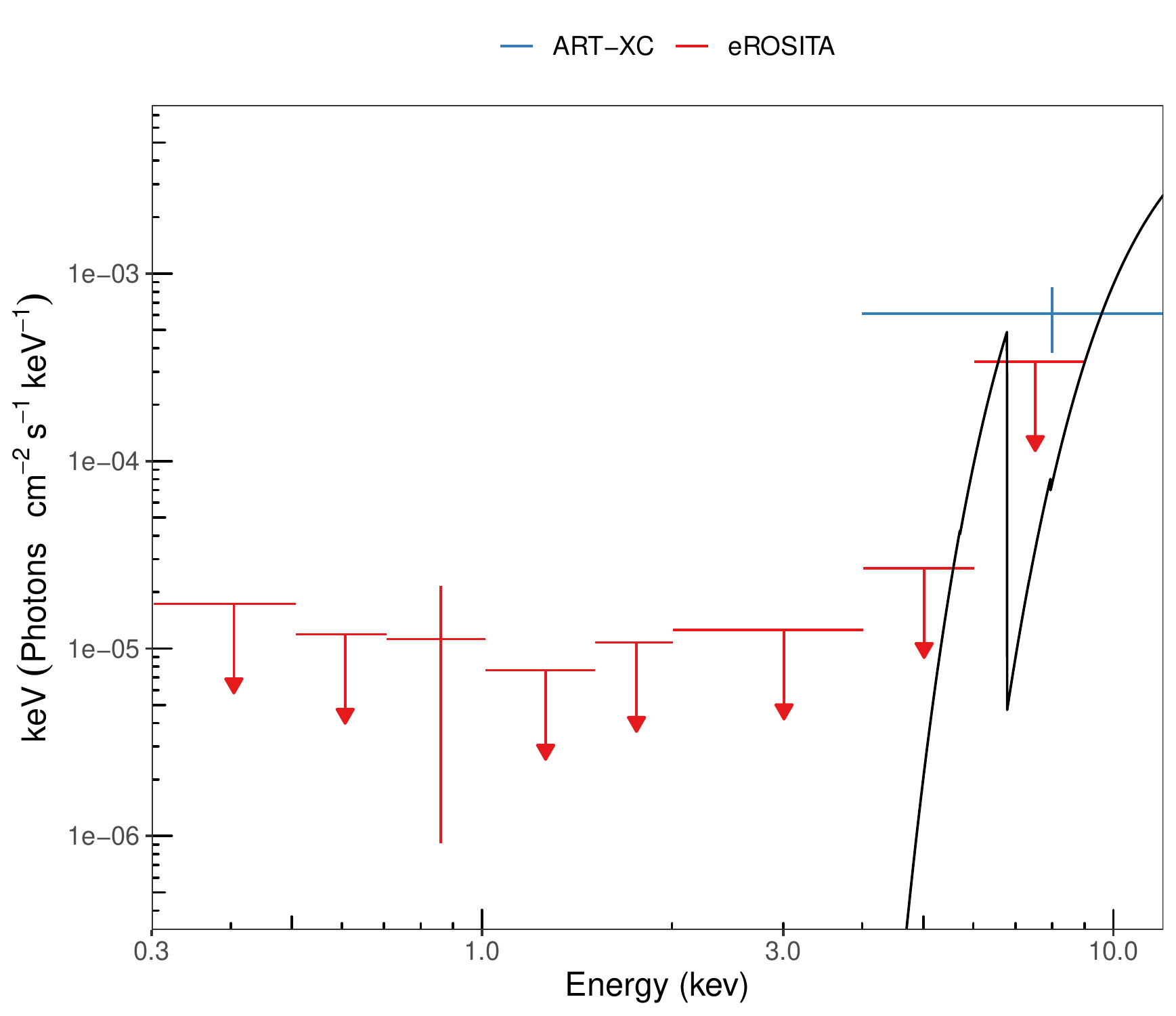}
  \end{floatrow}
  \vfill
  AZT-33IK
  \vfill
  \begin{floatrow}
    \smfigure{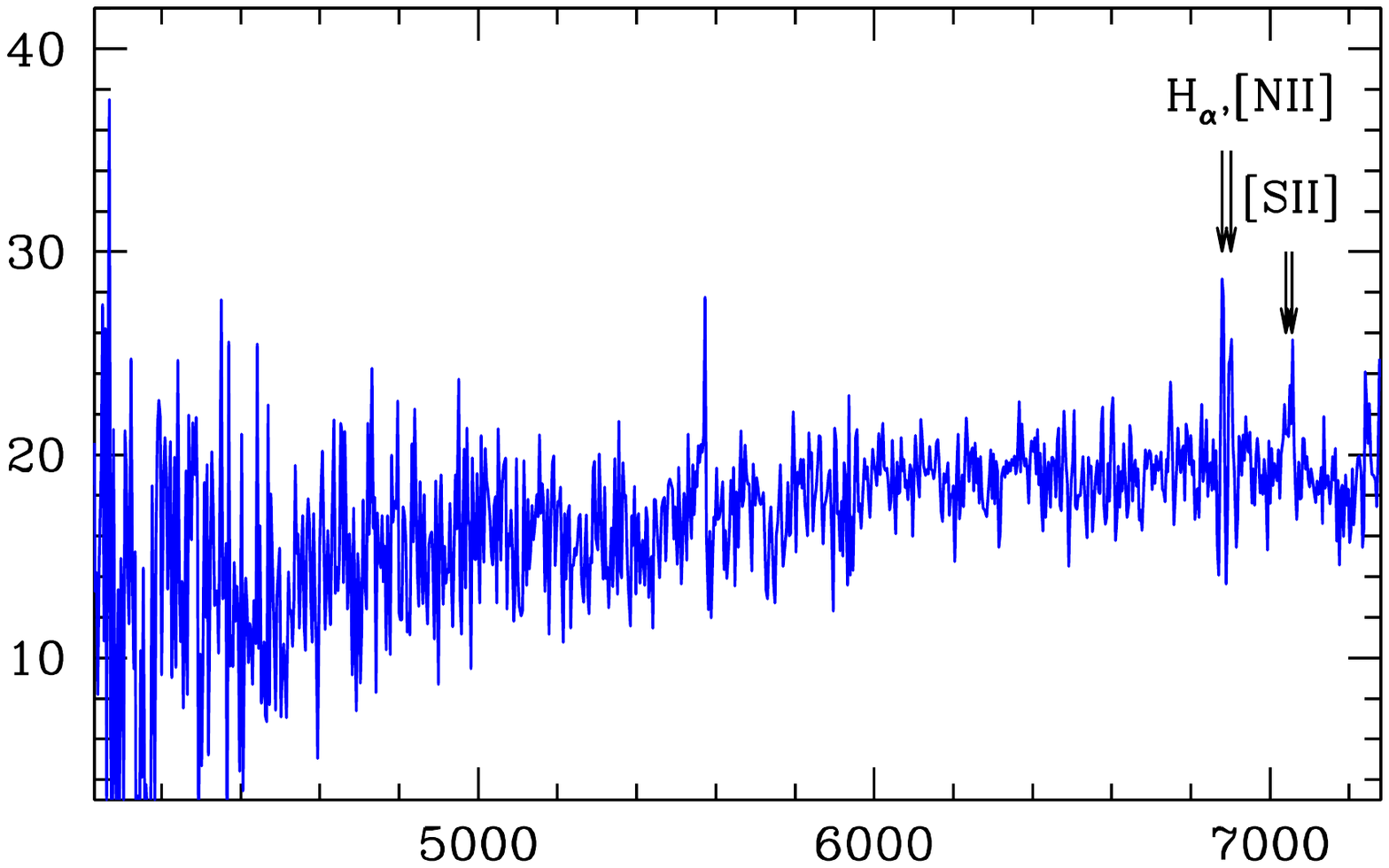}{$\lambda$, \AA}{Flux, $10^{-17}$\,erg\,s$^{-1}$\,cm$^{-2}$}
  \end{floatrow}
  \caption{Top left: the pointing picture for the source SRGA\,J$005751.0\!+\!210846$ taken from the \ps1\ survey \citep{ps1}. The arrow indicates the object for which an optical spectrum was taken; the blue circumference marks the \art\ position error circle of the source $30\arcsec$ in radius. Top right: the X-ray spectrum from the \art\ (red) and \ero\ (blue) data and the best-fit model (a power law with absorption) (black line). The arrows indicate the upper limits. Bottom: the optical spectrum with an indication of some emission and absorption lines.}
  \label{fig:spec0057}
\end{figure*}

\begin{table*}
  \caption{Spectral features of the SRGA\,J$005751.0\!+\!210846$ = LEDA 1643776} 
  \label{tab:j0057}
  \vskip 2mm
  \renewcommand{\arraystretch}{1.1}
  \renewcommand{\tabcolsep}{0.35cm}
  \centering
  \footnotesize
  \begin{tabular}{lcccc}
    \noalign{\doubleline}
    Line & Wavelength, \AA & Flux, $10^{-16}$\,erg\,s$^{-1}$\,cm$^{-2}$ & Eq. Width $^1$, \AA & $FWHM$, km\,s$^{-1}$\\
    \noalign{\vskip 3pt\hrule\vskip 5pt}
    H${\alpha}$ & 6880 & $9.9 \pm 6.0$ & $-5.0\pm 3.0$ & $(3.6\pm 0.3)\times 10^2$\\
    N\,II$\lambda$6584 & 6901 & $8.8\pm 5.0$ & $-4.4\pm 2.5$ & $(3.8\pm 0.2)\times 10^2$\\
    
    \noalign{\vskip 3pt\hrule\vskip 5pt}
  \end{tabular}
  
  \begin{flushleft}
  $^1$ Negative values correspond to emission lines.
  \end{flushleft}
\end{table*}

\section{RESULTS OF OBSERVATIONS}

Below we present the details of our optical and X-ray observations and the results obtained for each
object from the sample.

\subsection{\bf SRGA\,J005751.0\!+\!210846}
This X-ray source was discovered in the 4--12 keV band by the \art\ telescope of the \srg\ observatory and, at the same time, was not detected in softer X-rays by the \ero\ telescope of the same observatory.

A probable optical counterpart of the X-ray source is the galaxy LEDA~1643776 that falls into the \art\ position error circle (Fig.~\ref{fig:spec0057}). The galaxy is oriented edge-on to the observer. Previously, a spectrum has already been obtained for it during the Sloan Digital Sky Survey (release 12, \citealt{sdssdr12}), from which its redshift was measured ($z=0.04798\pm0.00002$). However, this spectrum does not allow the object to be reliably classified as an AGN.

Our optical observations of the object were carried out on October 22, 2020, at the AZT-33IK telescope with VPHG600G. Six spectral images with an exposure time of 300 s each were obtained; the total exposure time was 30 min.

Our spectrum of the galaxy is shown in Fig.~\ref{fig:spec0057}. Narrow H${\alpha}$, [N\,II]$\lambda$6584, and sulfur doublet emission lines are seen in it. The H${\beta}$, [O\,III]$\lambda$4959, and [O\,III]$\lambda$5007 lines are absent. The $2\sigma$ upper boundary of the intensity maximum in these lines is $5.5\times 10^{-17}$~erg\,s$^{-1}$\,cm$^{-2}$\,\AA$^{-1}$. Table~\ref{tab:j0057} presents the characteristics of the two brightest lines. From these two lines we measured the object’s redshift, $z=0.04795\pm0.00005$, consistent with the measured SDSS redshift within the error limits.

Because of the absence of the H${\beta}$ line and the [O\,III] doublet in the spectrum, it is impossible to establish the position of the galaxy LEDA~1643776 on the standard BPT diagram (\citealt{bpt}, see Fig.~\ref{chart:bpt}) and, consequently, its optical type. Nevertheless, a high X-ray luminosity ($\sim 5\times 10^{43}$~erg\,s$^{-1}$ in the 4--12 keV band from the \art\ data) of the object leaves no doubt that this is an AGN. The weakness of the observed lines probably stems from the fact that we observe the galaxy LEDA~1643776 edge-on, so that the optical radiation from the active nucleus (and, in particular, from the narrow-line region) is almost completely absorbed in the interstellar gas of the galaxy.

The non-detection by the \ero\ telescope in combination with the 4--12 keV flux measured by the \art\ telescope (Fig.~\ref{fig:spec0057}) allows a lower limit (at 90\% confidence) to be placed on the absorption column density: $\nh>10^{24}$~cm$^{-2}$. Much of this absorption may arise in the interstellar gas of the galaxy and not in the gas–dust torus surrounding the SMBH.

\subsection{\bf SRGA\,J$014157.0\!-\!032915$.}

This X-ray source was discovered in the 4--12 keV band by the \art\ telescope of the \srg\ observatory and, at the same time, was not detected in softer X-rays by the \ero\ telescope.

\begin{figure*}
  \centering
  \vfill
  SRGA\,J$014157.0\!-\!032915$
  \vfill
  \vskip 0.5cm
  \begin{floatrow}
    \includegraphics[width=0.35\columnwidth]{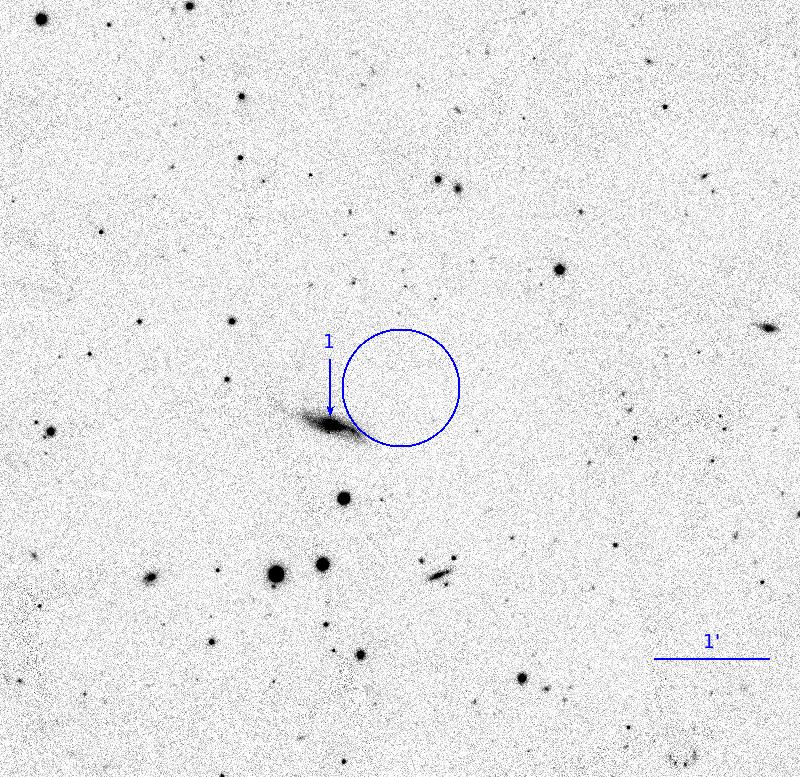}
    \includegraphics[width=0.35\columnwidth]{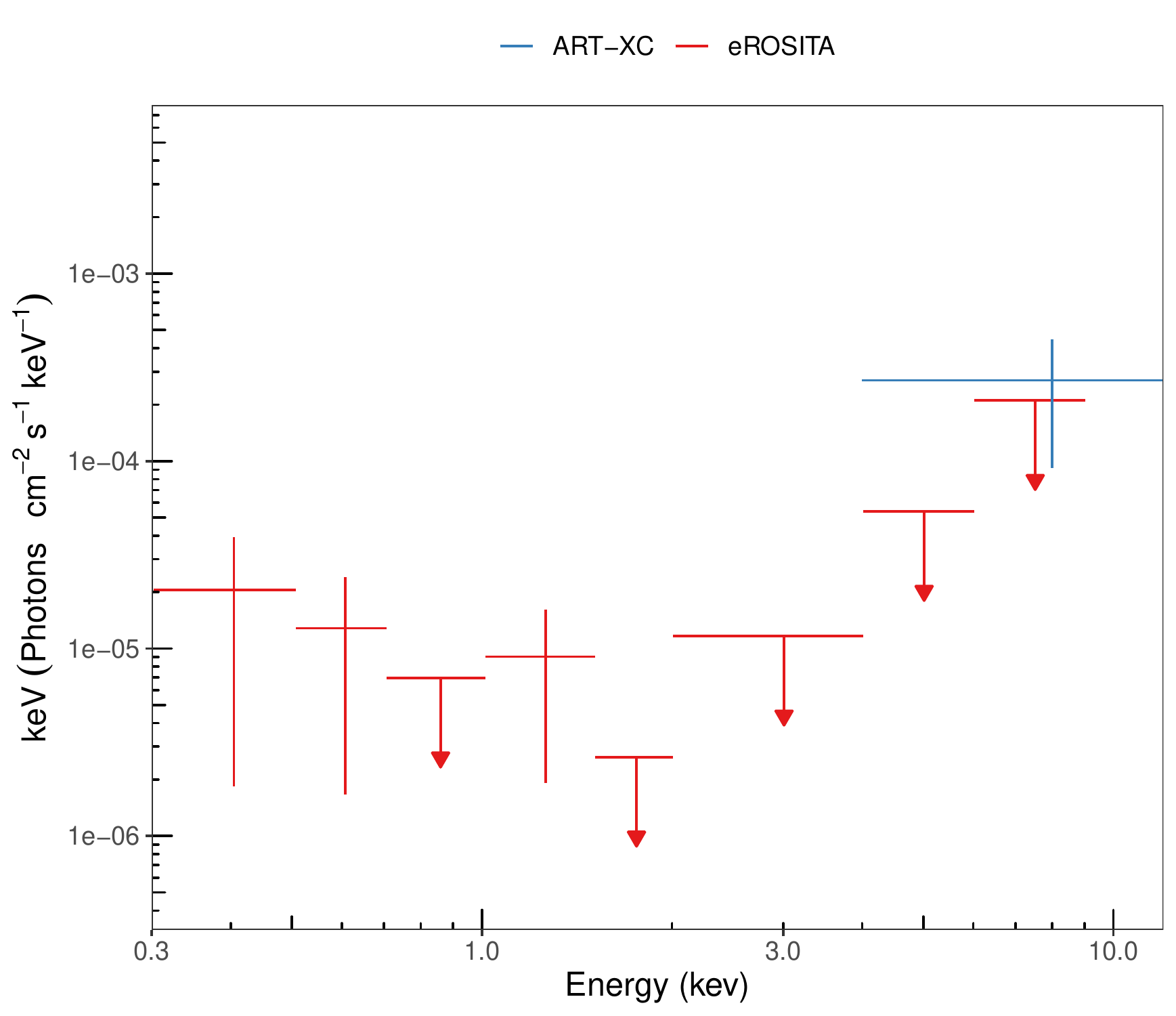}
  \end{floatrow}
  \vfill
  AZT-33IK
  \vfill
  \begin{floatrow}
    \smfigure{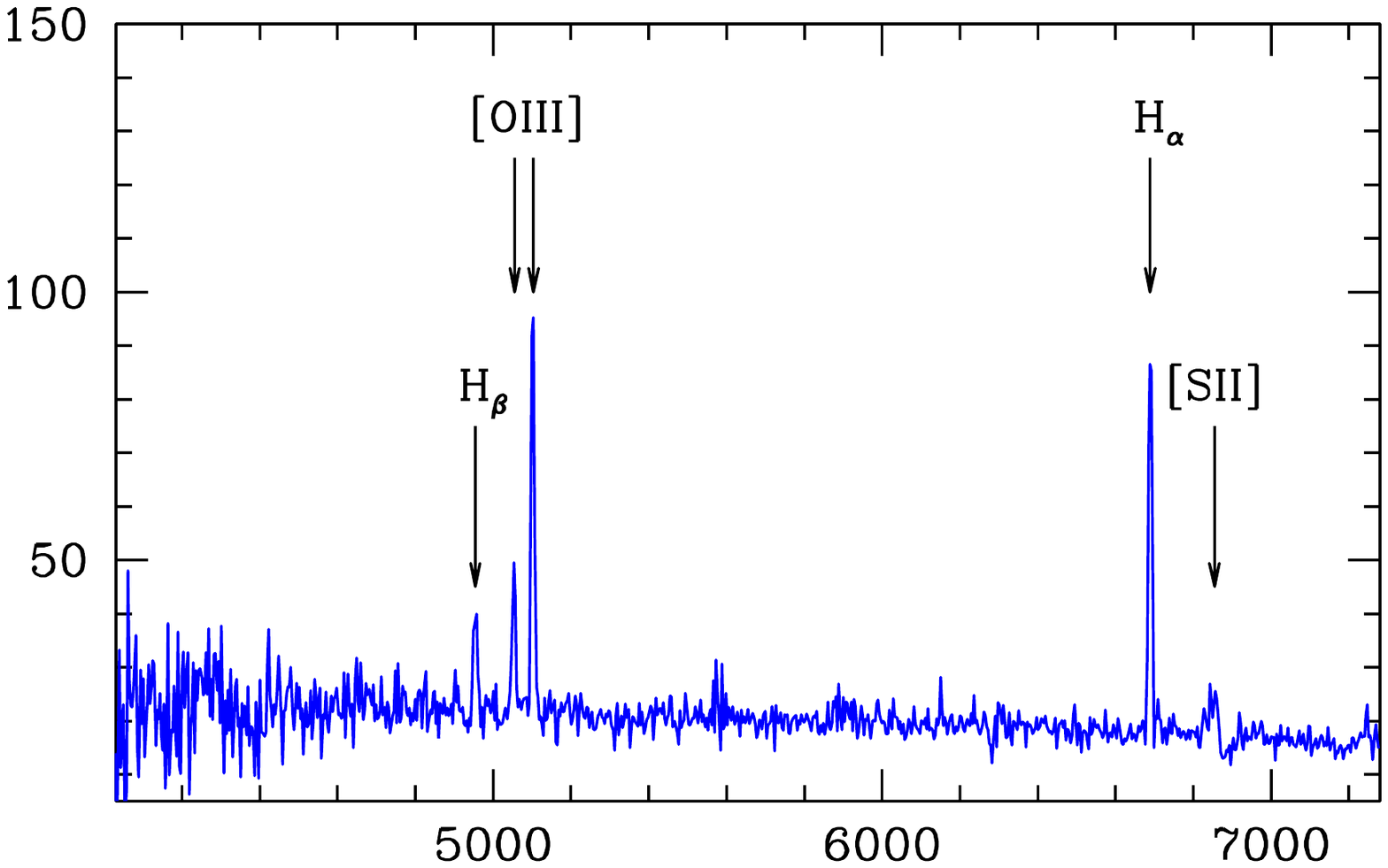}{$\lambda$, \AA}{Flux, $10^{-17}$ erg\,s$^{-1}$\,cm$^{-2}$}
  \end{floatrow}
  \caption{Same as Fig.~\ref{fig:spec0057}, but for SRGA\,J$014157.0\!-\!032915$. No spectral model is shown on the graph with the X-ray spectrum because of the large scatter in model parameters.}
  \label{fig:spec0141}
\end{figure*}

A probable optical counterpart is the galaxy LEDA\,1070544 (Fig.~\ref{fig:spec0141}). Although its center is at a distance of about $40\arcsec$ from the position of the X-ray source (Table \ref{tab:list}), such position errors can occur in the case of sources at the \art\ detection threshold.

\begin{table*}
  \caption{Spectral features of SRGA\,J$014157.0-032915$ = LEDA~1070544} 
  \label{tab:j0141}
  \vskip 2mm
  \renewcommand{\arraystretch}{1.1}
  \renewcommand{\tabcolsep}{0.35cm}
  \centering
  \footnotesize
  \begin{tabular}{lcccc}
    \noalign{\doubleline}
    Line & Wavelength, \AA& Flux, $10^{-16}$ erg\,s$^{-1}$\,cm$^{-2}$& Eq. Width, \AA& $FWHM$, km\,s$^{-1}$\\
    \noalign{\vskip 3pt\hrule\vskip 5pt}
    H${\beta}$ & 4954 & $27\pm 5$ & $-17.8\pm 3.0$ & $(6.0\pm 0.7)\times 10^2$\\
    O\,III$\lambda$4960 &	5054 & $31\pm 3$ & $-14.5\pm 1.5$ & $(5.0\pm 0.7)\times 10^2$\\
    O\,III$\lambda$5007 & 5102 & $84\pm 9$ & $-38\pm 4$ & $(4.9\pm 0.7)\times 10^2$\\
    H${\alpha}$ & 6689 & $88\pm 8$ & $-50\pm 5$ & $(3.9\pm 0.5)\times 10^2$\\
    N\,II$\lambda$6584 & 6709 & $<12$ & $>-7.3$ & --\\ 

  \noalign{\vskip 3pt\hrule\vskip 5pt}
  \end{tabular}
\end{table*}

Our optical observations were carried out on October 13, 2020, at the AZT-33IK telescope using VPHG600G. Three spectral images with an exposure time of 600~s each were obtained; the total exposure time was 30~min.

The H${\beta}$, [O\,III]$\lambda$4959, [O\,III]$\lambda$5007, H${\alpha}$, and [S\,II] doublet emission lines are seen in our spectrum (Fig.~\ref{fig:spec0141}). The [N\,II]$\lambda$6584 line is difficult to separate from the H${\alpha}$ line. Table~\ref{tab:j0141} gives the characteristics of the emission lines. They are all narrow. The redshift determined from four lines is $z=0.01878\pm0.00003$.

The absence of broad lines in the spectrum and a fairly high X-ray luminosity ($\sim 3\times 10^{42}$\,erg\,s$^{-1}$ in the 4--12 keV band from the \art\ data) suggest that this is a Seyfert 2 galaxy. However, according to the measured line flux ratios log([O\,III]$\lambda 5007/$H${\beta}) = 0.49\pm 0.09$ and log([N\,II]$\lambda 6584/$H${\alpha})<-0.86$, the object lies in the region of star-forming galaxies on the BPT diagram (Fig.~\ref{chart:bpt}), though near (within three standard deviations) the region of Seyfert galaxies. Most likely, we are dealing with a galaxy in which, apart from SMBH activity, there is active star formation.

The non-detection by the \ero\ telescope in combination with the 4--12 keV flux measured by the \art\ telescope (Fig.~\ref{fig:spec0057}) allows a lower limit to be placed on the absorption column density: $\nh>3\times 10^{23}$\,cm$^{-2}$. However, this limit cannot yet be deemed reliable, because it was obtained only at 68\% confidence. To make sure that there is a strong absorption in this object, it is necessary to take an X-ray spectrum with a significantly larger number of photons.

\subsection{\bf SRGA\,J043209.6\!+\!354917}
This X-ray source is first mentioned under the name 1WGA\,J$0432.1\!+\!3549$ in the catalogue of sources discovered in the soft X-ray energy band during the \rosat\ pointed observations \citep{wgacat}. It is also present in the catalogue of sources detected during the \xmm\ slew observations \citep{xmmsl2}. However, the nature of this object so far has remained unknown. The source was reliably detected by both \art\ and \ero\ telescopes of the \srg\ observatory.

\begin{figure*}
  \centering
  \vfill
  SRGA\,J$043209.6\!+\!354917$
  \vfill
  \vskip 0.5cm
  \begin{floatrow}  
    \includegraphics[width=0.40\columnwidth]{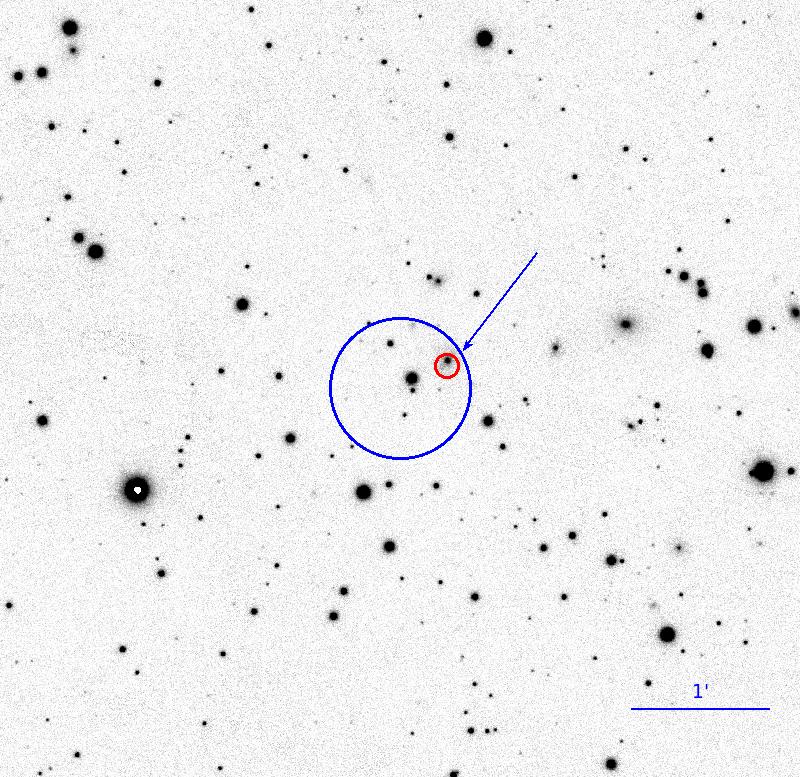}
    \includegraphics[width=0.45\columnwidth]{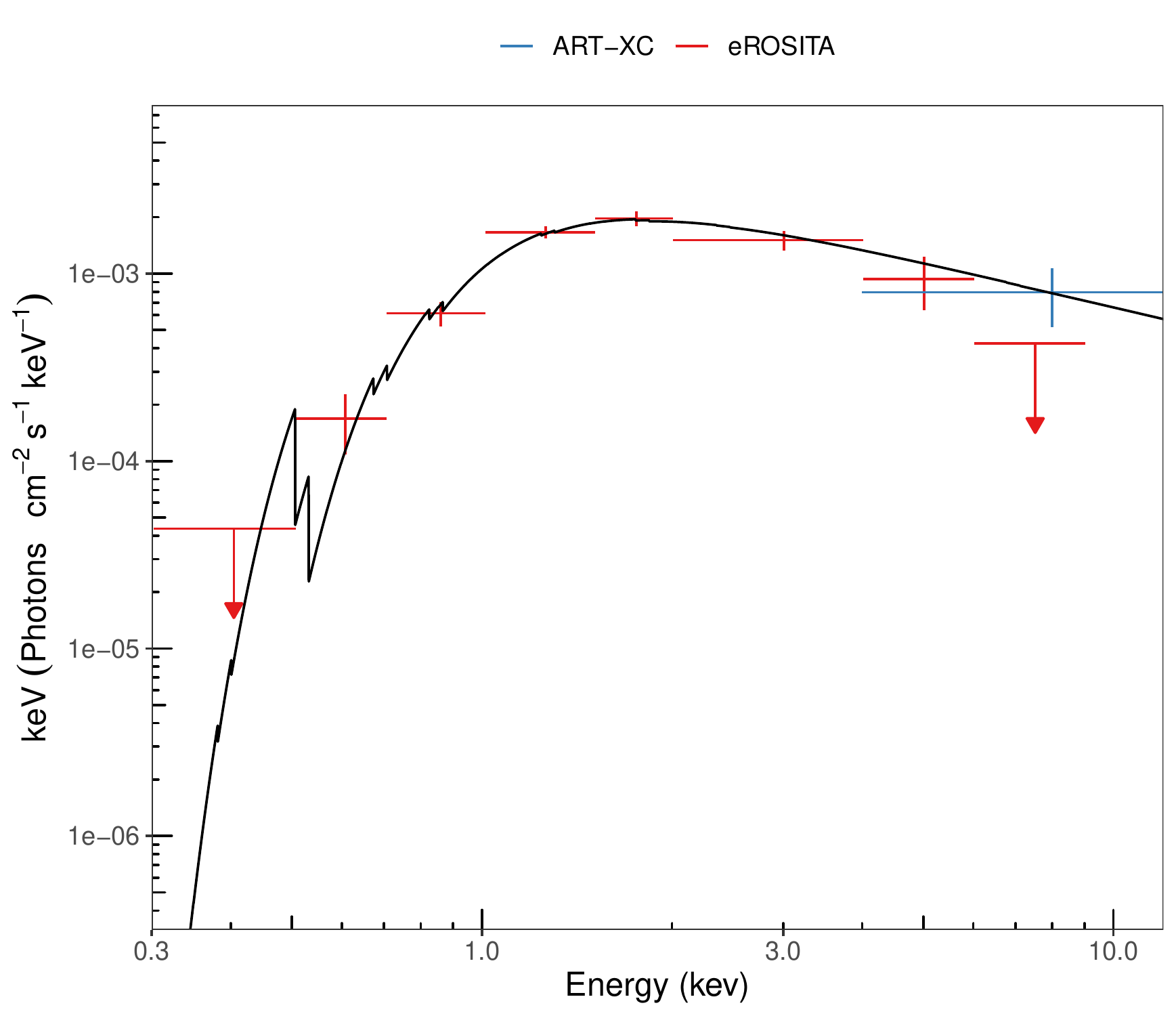} 
  \end{floatrow}
  \vfill
  RTT-150
  \vfill
  \begin{floatrow}
  \smfigure{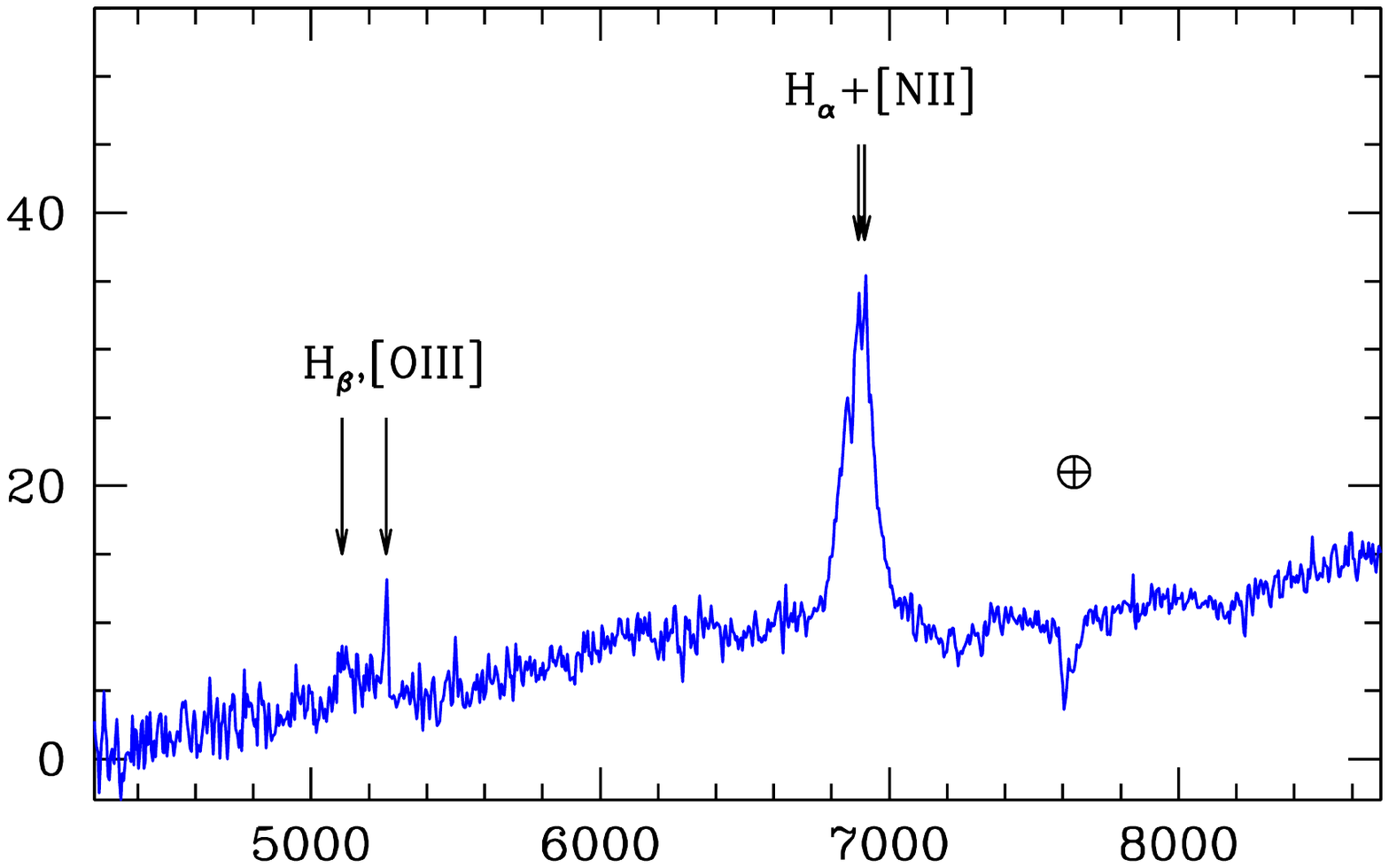}{$\lambda$, \AA}{Flux, $10^{-17}$ erg\,s$^{-1}$\,cm$^{-2}$}
  \end{floatrow}
  \caption{Same as Fig.~\ref{fig:spec0057}, but for SRGA\,J$043209.6\!+\!354917$. In the pointing picture the blue and red circumferences indicate the \art\ and \ero\ position error circles, respectively.}
  \label{fig:spec0432}
\end{figure*}

The X-ray source is reliably identified with the galaxy 2MASX\,J$04320796\!+\!3549287$ = WISEA\,J$043207.95\!+\!354928.8$ (Fig.~\ref{fig:spec0432}), whose colors in the near infrared ($W1-W2=0.68$) \citep{wright10} point to the presence of an active nucleus.

\begin{table*}
  \caption{Spectral features of SRGA\,J$043209.6\!+\!354917$ = 2MASX\!J$04320796\!+\!3549287$} 
  \label{tab:j0432}
  \vskip 2mm
  \renewcommand{\arraystretch}{1.1}
  \renewcommand{\tabcolsep}{0.35cm}
  \centering
  \footnotesize
  \begin{tabular}{lcccc}
    \noalign{\doubleline}
    Line & Wavelength, \AA& Flux, $10^{-16}$ erg~\,s$^{-1}$\,cm$^{-2}$& Eq. Width, \AA& $FWHM$, km\,s$^{-1}$\\
    \noalign{\vskip 3pt\hrule\vskip 5pt}
    H${\beta}$, narrow & 5112 & $<2$ & $>-4.3$ & --\\
    H${\beta}$, broad & 5112 & $28 \pm 7$ & $-64\pm 16$ & $(5.8\pm 0.6)\times 10^3$\\
    O\,III$\lambda$4960 & -- & $<3$ & $>-6.5$ & --\\
    O\,III$\lambda$5007 & 5260 & $12\pm 2$ & $-27\pm 5$ & $(6.4\pm 1.2)\times 10^2$\\
    N\,II$\lambda$6548 & 6854 & $2\pm 2$ & $-2\pm 2$ & $(5.8\pm 0.9)\times 10^2$\\
    H${\alpha}$, narrow & 6893 & $9\pm 2$ & $-9\pm 2$ & $(5.8\pm 0.9)\times 10^2$\\
    H${\alpha}$, broad & 6893 & $282\pm 8$ & $-278\pm 8$ & $(6.0\pm 0.2)\times 10^3$\\
    N\,II$\lambda$6584 & 6919 & $14\pm 2$ & $-14\pm 2$ & $(5.8\pm 0.9)\times 10^2$\\

  \noalign{\vskip 3pt\hrule\vskip 5pt}
  \end{tabular}
\end{table*}

Our optical observations were carried out on September 15, 2020, at RTT-150. Five spectral images with an exposure of 900~s each were obtained; the total exposure time was 75~min.

Balmer hydrogen and forbidden oxygen and nitrogen emission lines are seen in the object’s spectrum (Fig.~\ref{fig:spec0432}). Table~\ref{tab:j0432} gives the characteristics of the emission lines. The redshift was determined from three lines, H${\alpha}$, [O\,III]$\lambda$5007, and [N\,II]$\lambda$6584, and is $z=0.0506\pm 0.0010$. The H${\alpha}$ and H${\beta}$ lines have a broad component; only an upper limit can be placed on the flux in the narrow H${\beta}$ component. The object is classified from the ratios log([O\,III]$\lambda5007/$H${\beta})>0.77$ and log([N\,II]$\lambda6584/$H${\alpha})=0.19\pm 0.11$ (Fig.~\ref{chart:bpt}) and the fluxes in the broad and narrow H${\alpha}$ and H${\beta}$ components as a Seyfert 1 galaxy.
A slight absorption is detected in the object’s X- ray spectrum (Fig.~\ref{fig:spec0432}): $\nh\sim 3\times 10^{21}$~cm$^{-2}$.

\subsection{\bf SRGA\,J045049.8\!+\!301449}

This object was discovered in hard X-rays (the source SWIFT\,J0450.6\!+\!3015) by the BAT instrument of the Neil Gehrels \swift\ observatory \citep{swiftbat} and is present in the catalogue of point X-ray sources detected by the XRT telescope of the same observatory \citep{evan20}. However, its nature so far has remained unknown. The source was reliably detected by both \art\ and \ero\ telescopes of the \srg\ observatory.

\begin{figure*}
  \centering
  \vfill
  SRGA\,J$045049.8\!+\!301449$
  \vfill
  \vskip 2cm
  \begin{floatrow}
    \includegraphics[width=0.40\columnwidth]{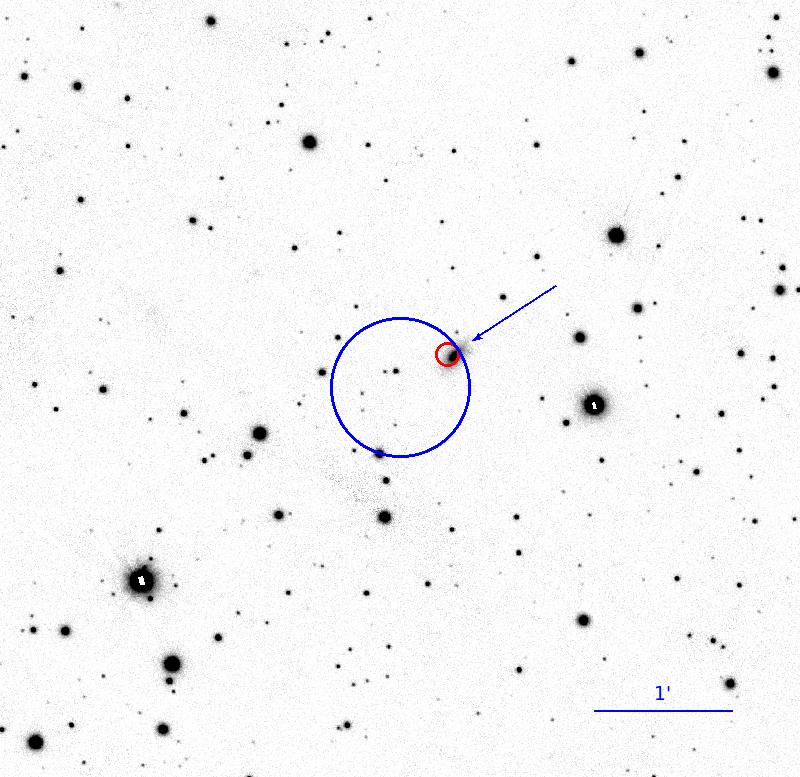}
    \includegraphics[width=0.40\columnwidth]{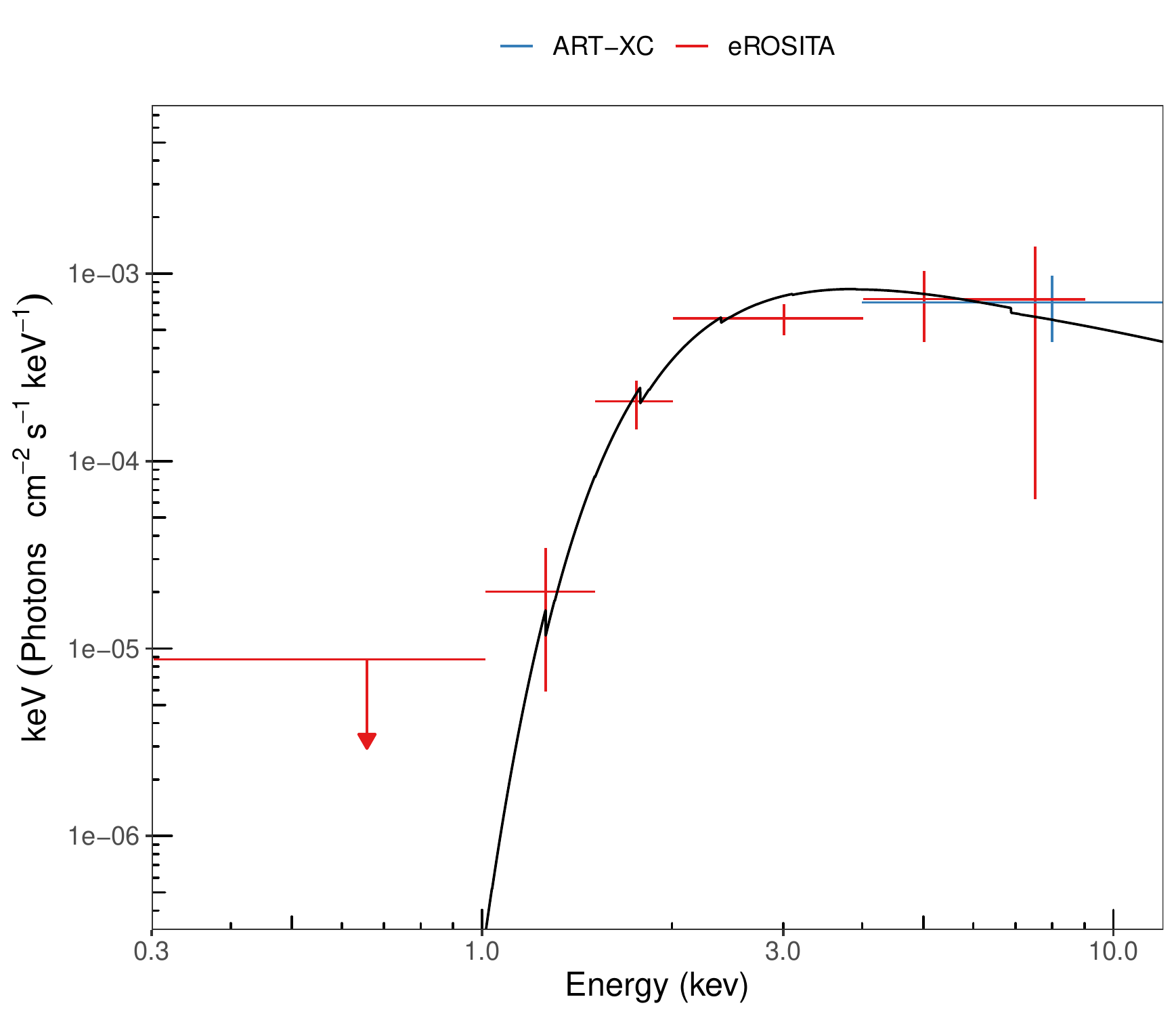}
  \end{floatrow}
  \vfill
  AZT-33IK
  \vfill
  \begin{floatrow}
  \smfigure{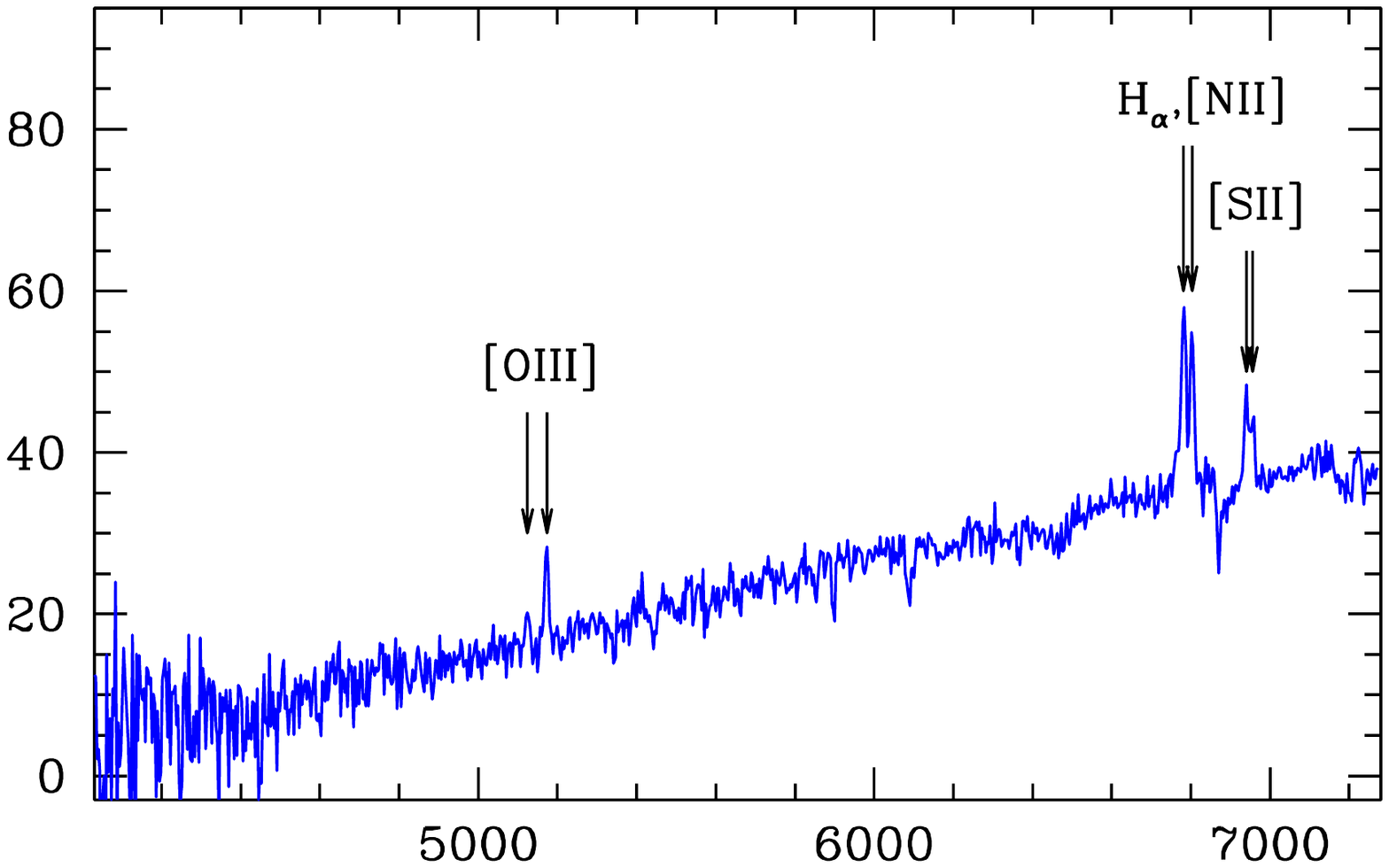}{$\lambda$, \AA}{Flux, $10^{-17}$ erg\,s$^{-1}$\,cm$^{-2}$}
  \end{floatrow}
  \bigskip

  \caption{Same as Fig.~\ref{fig:spec0432}, but for SRGA\,J$045049.8\!+\!301449$.}
  \label{fig:spec0450}
\end{figure*}

The X-ray source is reliably identified (Fig.~\ref{fig:spec0450}) with the galaxy LEDA\,1896296 = WISEA\,J045048.00\!+\!301502.8 ($W1-W2=0.38$).

\begin{table*}
  \caption{Spectral features of SRGA\,J$045049.8\!+\!301449$ = LEDA~1896296} 
  \label{tab:j0450}
  \vskip 2mm
  \renewcommand{\arraystretch}{1.1}
  \renewcommand{\tabcolsep}{0.35cm}
  \centering
  \footnotesize
  \begin{tabular}{lcccc}
    \noalign{\doubleline}
    Line & Wavelength, \AA& Flux, $10^{-16}$ erg\,s$^{-1}$\,cm$^{-2}$& Eq. Width, \AA& $FWHM$, km\,s$^{-1}$\\
    \noalign{\vskip 3pt\hrule\vskip 5pt}
    H${\beta}$ & 5037 & $<2$ & $>-1.3$ & --\\
    O\,III$\lambda$4960 & 5124 & $5.9\pm 0.8$ & $-3.7\pm 0.5$ & $(6.5\pm 0.6)\times 10^2$\\
    O\,III$\lambda$5007 & 5173 & $17\pm 2$ & $-10.3\pm 1.2$ & $(6.8\pm 0.6)\times 10^2$\\
    H${\alpha}$, narrow & 6781 & $24\pm 8$ & $-7\pm 2$ & $(4.7\pm 0.5)\times 10^2$\\
    H${\alpha}$, broad & 6781 & $25\pm 4$ & $-7.1\pm 1.2$ & $(2.8\pm 0.4)\times 10^3$\\
    N\,II$\lambda$6584 & 6803 & $22\pm 3$ & $-6.7\pm 0.9$ & $(4.8\pm 0.5)\times 10^2$\\
    SII$\lambda$6718 & 6940 & $15\pm 2$ & $-4.2\pm 0.6$ & $(5.2\pm 0.8)\times 10^2$\\
    SII$\lambda$6732 & 6956 & $10\pm 2$ & $-2.7\pm 0.6$ & $(4.8\pm 0.8)\times 10^2$\\

  \noalign{\vskip 3pt\hrule\vskip 5pt}
  \end{tabular}
\end{table*}

Our optical observations were carried out on October 22, 2020, at the AZT-33IK telescope using VPHG600G. Four spectral images with an exposure time of 600 s each were obtained near the object’s culmination; the total exposure time was 40 min.

The [O\,III]$\lambda$4960, $\lambda$5007, H${\alpha}$, [N\,II]$\lambda$6584, and sulfur doublet emission lines are seen in the object’s spectrum (Fig.~\ref{fig:spec0450}). The H${\beta}$ line is unseen. The upper limit on the ratio log([O\,III]$\lambda 5007/$H${\beta})>0.92$. The upper limit on the ratio log([N\,II]$\lambda 6584/$H${\alpha})=-0.04\pm 0.16$. All these lines are narrow, except H${\alpha}$ in which a broad component can be distinguished.

The characteristics of the emission lines are given in Table~\ref{tab:j0450}. The object’s redshift was measured from six emission lines: $z=0.03308 \pm 0.00004$. From the position on the BPT diagram (Fig. ~\ref{chart:bpt}) and the presence of a broad component only in the Balmer H${\alpha}$ line, the object can be classified as a Seyfert 1.9 galaxy.

An appreciable absorption is detected in the object’s X-ray spectrum (Fig.~\ref{fig:spec0450}): $\nh\sim 4\times 10^{22}$~cm$^{-2}$.

\subsection{\bf SRGA\,J152102.3\!+\!320418}

This X-ray source is present in the catalogue of point X-ray sources detected by the XRT telescope of the Neil Gehrels \swift\ observatory \citep{evan20}, but its nature so far has remained unknown. The source was reliably detected by both \art\ and \ero\ telescopes of the \srg\ observatory.

\begin{figure*}
  \centering
  \vfill
  SRGA\,J$152102.3\!+\!320418$
  \vfill
  \vskip 0.5cm
  \begin{floatrow}  
    \includegraphics[width=0.40\columnwidth]{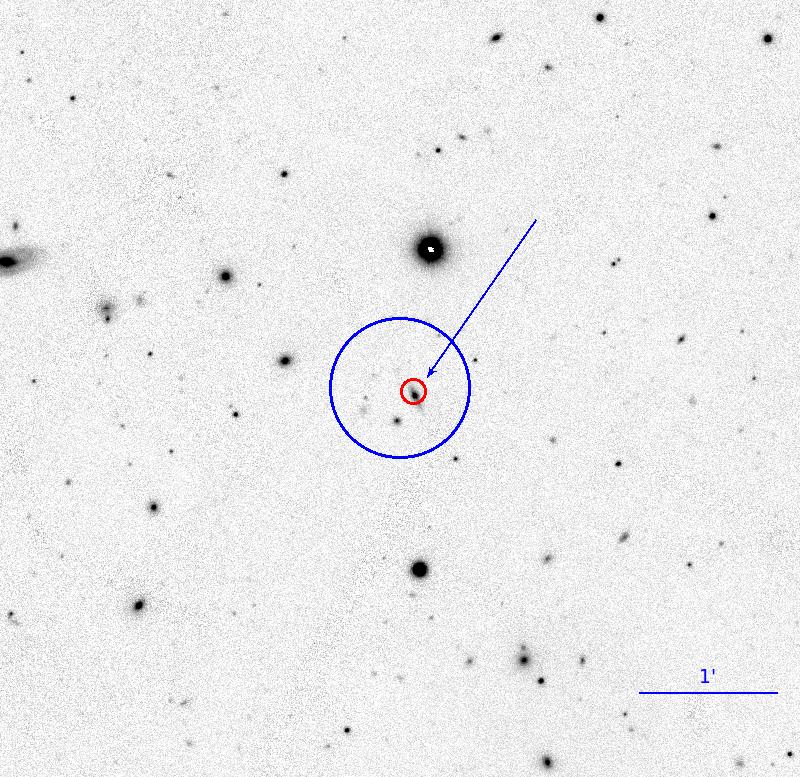}
    \includegraphics[width=0.45\columnwidth]{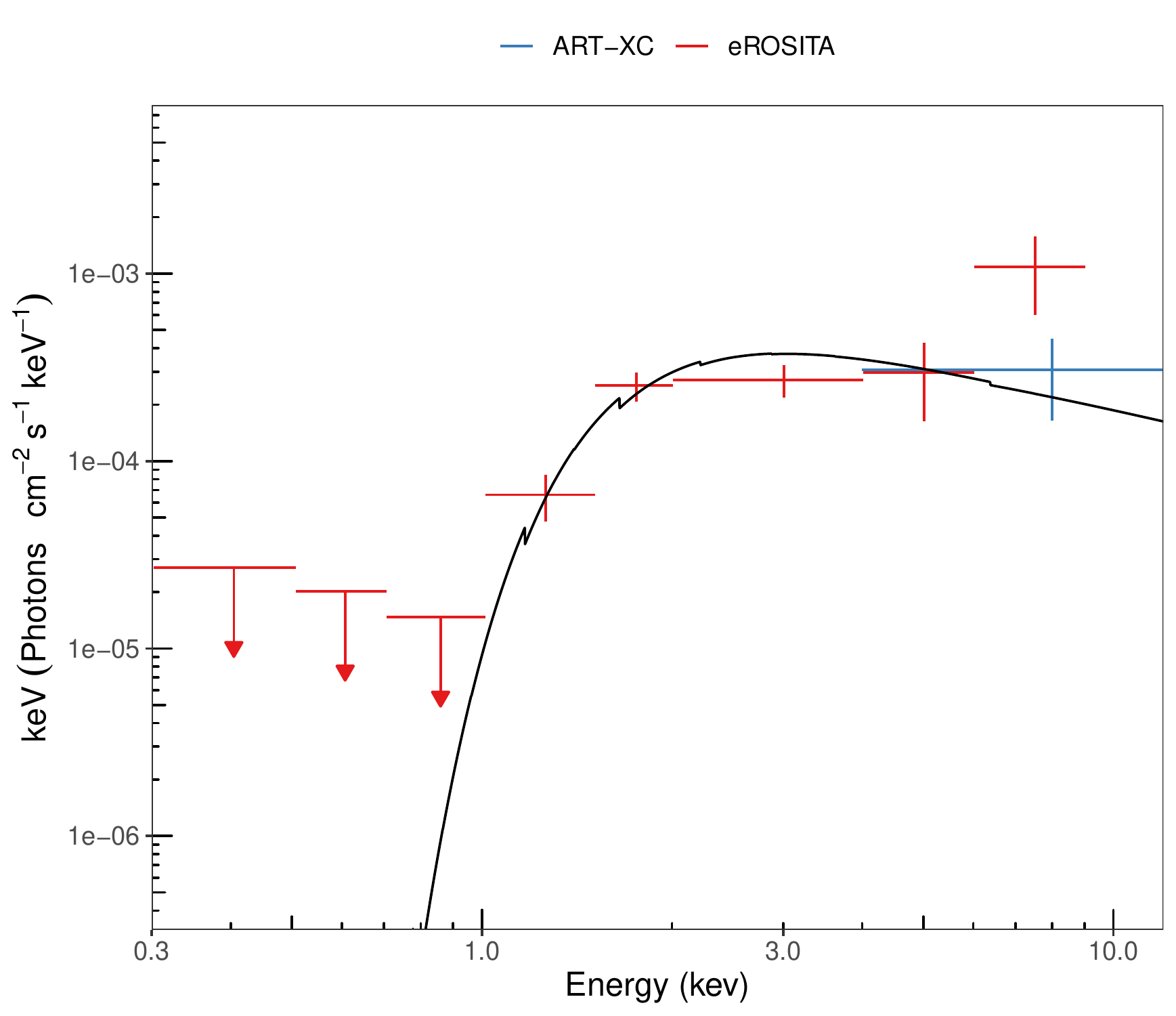}
  \end{floatrow}
  \vfill
  AZT-33IK
  \vfill
  \begin{floatrow}
  \smfiguresmall{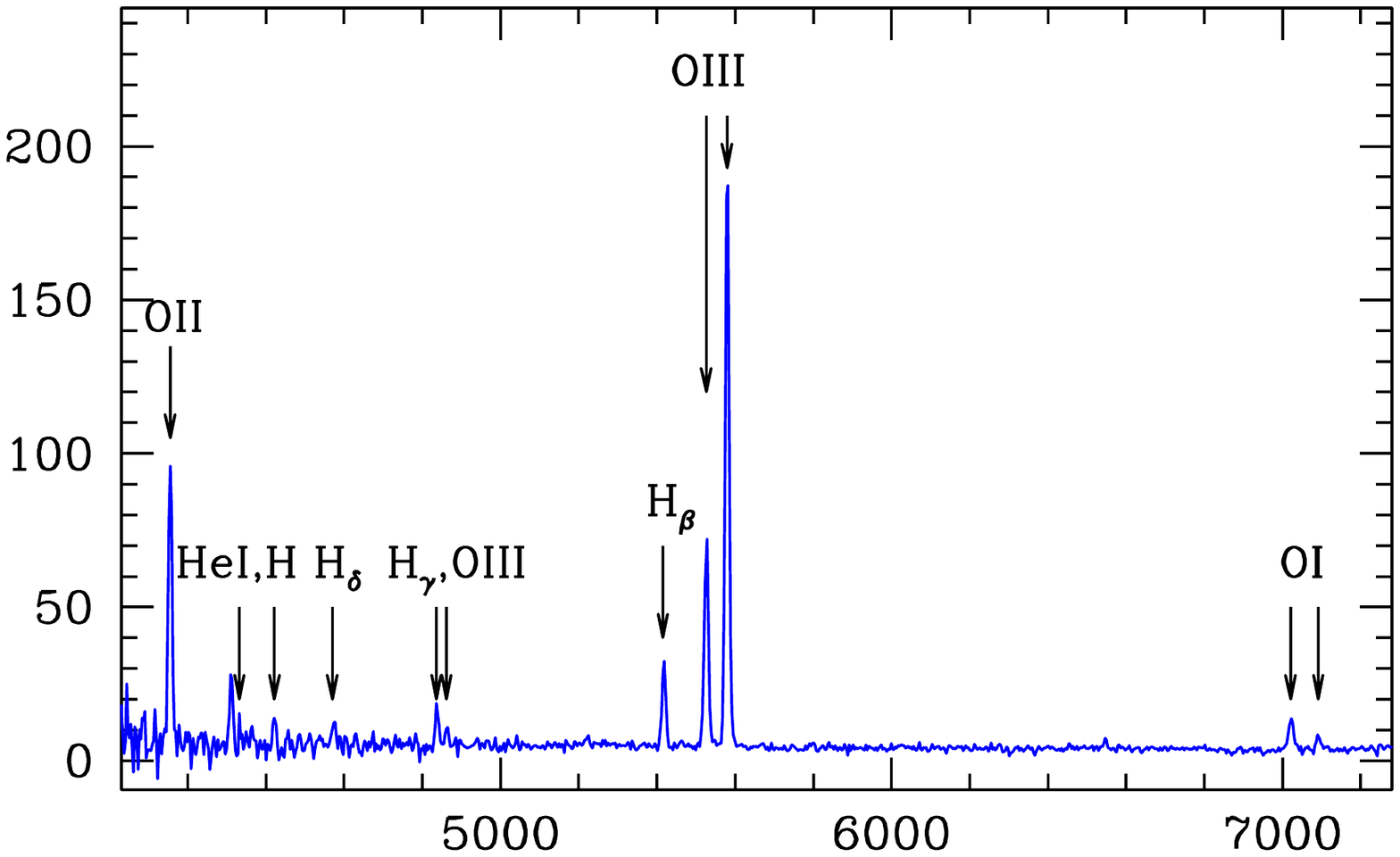}{$\lambda$, \AA}{Flux, $10^{-17}$ erg\,s$^{-1}$\,cm$^{-2}$}
  \smfiguresmall{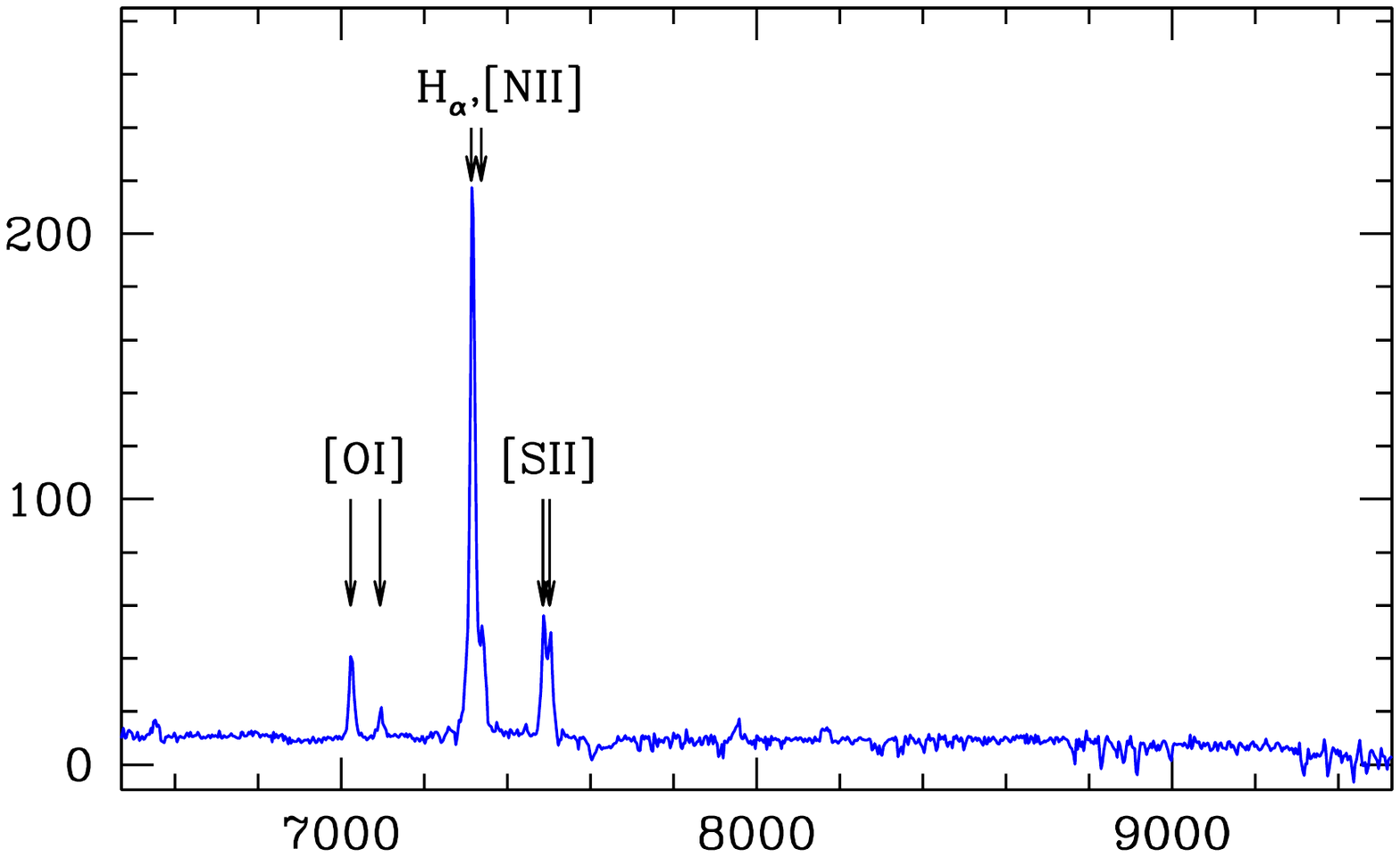}{$\lambda$, \AA}{Flux, $10^{-17}$ erg\,s$^{-1}$\,cm$^{-2}$}
  \end{floatrow}
  \bigskip

  \caption{Same as Fig.~\ref{fig:spec0432}, but for SRGA\,J$152102.3\!+\!320418$. The optical spectrum is shown on the two lower panels: the spectrum taken in VPGH600G (left) and VPHG600R (right).}
  \label{fig:spec1521}
\end{figure*}

The X-ray source is reliably identified (Fig.~\ref{fig:spec1521}) with the galaxy (SDSS data) WISEA\,J152101.83\!+\!320414.6, whose infrared color ($W1-W2=1.20$) points to the possible presence of an active nucleus.

\begin{table*}
  \caption{Spectral features of SRGA\,J$152102.3\!+\!320418$ = WISEA\,J$152101.83\!+\!320414.6$}
  \label{tab:j1521}
  \vskip 2mm
  \renewcommand{\arraystretch}{1.1}
  \renewcommand{\tabcolsep}{0.35cm}
  \centering
  \footnotesize
  \begin{tabular}{lcccc}
    \noalign{\doubleline}
    Line & Wavelength, \AA& Flux, $10^{-16}$~erg\,s$^{-1}$\,cm$^{-2}$& Eq. Width, \AA& $FWHM$, km\,s$^{-1}$\\
    \noalign{\vskip 3pt\hrule\vskip 5pt}
    OII$\lambda$3729 & 4155 & $112\pm 7$ & $(-1.8^{+1.7}_{-0.8})\times 10^2$ & $(6.8\pm 0.8)\times 10^2$\\
    HeI$\lambda$3889 & 4312 & $33\pm 7$ & $-9^{+9}_{-4} $ & $(6.2\pm 0.8)\times 10^2$\\
    H${\delta}$ & 4573 & $<11$ & $>-27$	& --\\
    H${\gamma}$ & 4837 & $14\pm 2$ & $-31^{+12}_{-6}$ & $(6.7\pm 0.7)\times 10^2$\\
    O\,III$\lambda$4364 & 4863 & $7.1\pm 1.5$ & $-15.4^{+1.2}_{-2.7}$ & $(6.3\pm 0.7)\times 10^2$\\
    H${\beta}$ & 5418 & $30\pm 3$ & $-67^{+24}_{-13}$ & $(5.7\pm 0.6)\times 10^2$\\
    O\,III$\lambda$4960 & 5527 & $75\pm 3$ & $(-1.23^{+0.45}_{-0.11})\times 10^2$ & $(6.1\pm 0.6)\times 10^2$\\
    O\,III$\lambda$5007 & 5580 & $(2.3\pm 0.1)\cdot 10^2$ & $(-3.6\pm 0.4)\times 10^2$ & $(6.3\pm 0.6)\times 10^2$\\
    OI$\lambda$6302 & 7022 & $14.9\pm 0.7$ & $-41^{+8}_{-7}$ & $(6.1\pm 0.5)\times 10^2$\\
    OI$\lambda$6365 & 7086 & $<7$ & $>-7.6$ & --\\
    N\,II$\lambda$6548 & 7298 & $<20$ & $>-22$ & --\\
    H${\alpha}$ & 7316 & $(3.0\pm 0.1)\cdot 10^2$ & $(-2.4^{+0.8}_{-0.7})\times 10^2$ & $(5.6\pm 0.4)\times 10^2$\\
    N\,II$\lambda$6584 & 7339 & $73\pm 4$ & $-56^{+29}_{-28}$ & $(7.2\pm 0.8)\times 10^2$\\		
    SII$\lambda$6718 & 7487 & $62\pm 7$ & $-58\pm 8$ & $(5.9\pm 0.4)\times 10^2$\\
    SII$\lambda$6732 & 7504 & $53\pm 7$ & $-50\pm 8$ & $(5.8\pm 0.4)\times 10^2$\\

  \noalign{\vskip 3pt\hrule\vskip 5pt}
  \end{tabular}
\end{table*}

Our optical observations were carried out on February 27 and April 24, 2020, at the AZT-33IK telescope. Five spectral images with an exposure time of 600 s each were obtained on February 27, 2020, the total exposure time was 50~min; on April 24, 2020, we obtained two spectral images with an exposure time of 1200 s each in VPHG600G and three spectral images with an exposure time of 1200 s each in VPHG600R, the total exposure time was 100 min.

Fourteen narrow hydrogen, oxygen, nitrogen, sulfur, and helium emission lines are seen in our spectrum (Fig.~\ref{fig:spec1521}). Information on these lines is collected in Table~\ref{tab:j1521}. The redshift of the galaxy determined from these 14 lines is $z=0.11425\pm 0.00031$. The ratios log([N\,II]$\lambda 6584/$H${\alpha})=-0.61\pm 0.03$ and log([O\,III]$\lambda 5007/$H${\beta})=0.88\pm 0.05$. From the position on the BPT diagram (Fig.~\ref{chart:bpt}) and the absence of broad lines, the object is classified as a Seyfert 2 galaxy.

An appreciable absorption is detected in the object’s X-ray spectrum (Fig.~\ref{fig:spec1521}): $\nh\sim 2.5\times 10^{22}$~cm$^{-2}$.

\subsection{\bf SRGA\,J200431.6\!+\!610211}
This X-ray source was discovered during the \rosat\ all-sky survey: 2RXS\,J$200433.8\!+\!610235$ \citep{2rxs}. However, its nature so far has remained unknown. The source was detected by both \art\ and \ero\ telescopes of the \srg\ observatory.

\begin{figure*}
  \centering
  \vfill
  SRGA\,J$200431.6\!+\!610211$
  \vfill
  \vskip 0.5cm
  \begin{floatrow}  
    \includegraphics[width=0.40\columnwidth]{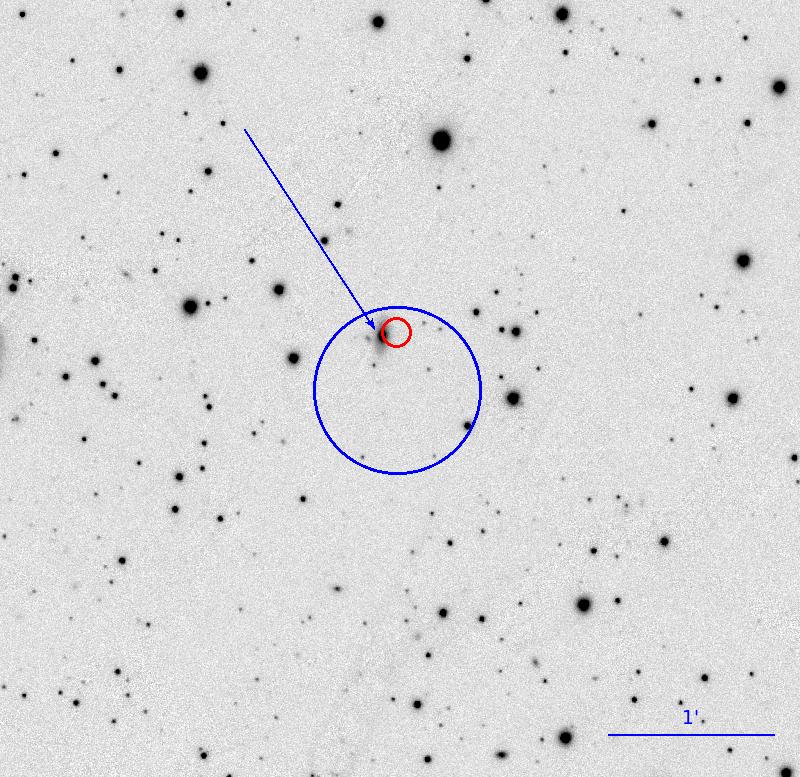}
    \includegraphics[width=0.45\columnwidth]{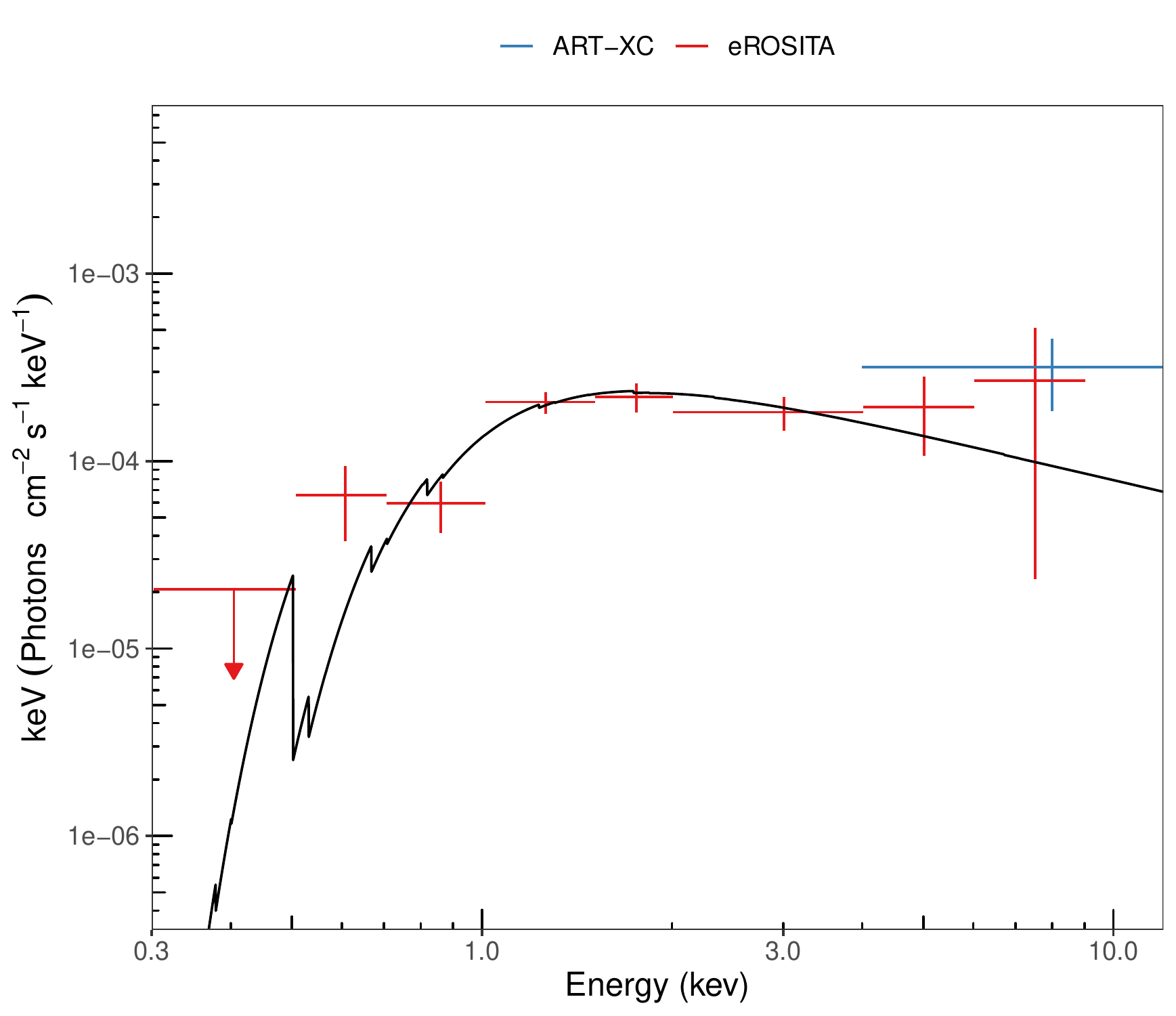}
  \end{floatrow}
  \vfill
  AZT-33IK
  \vfill
  \begin{floatrow}
  \smfigure{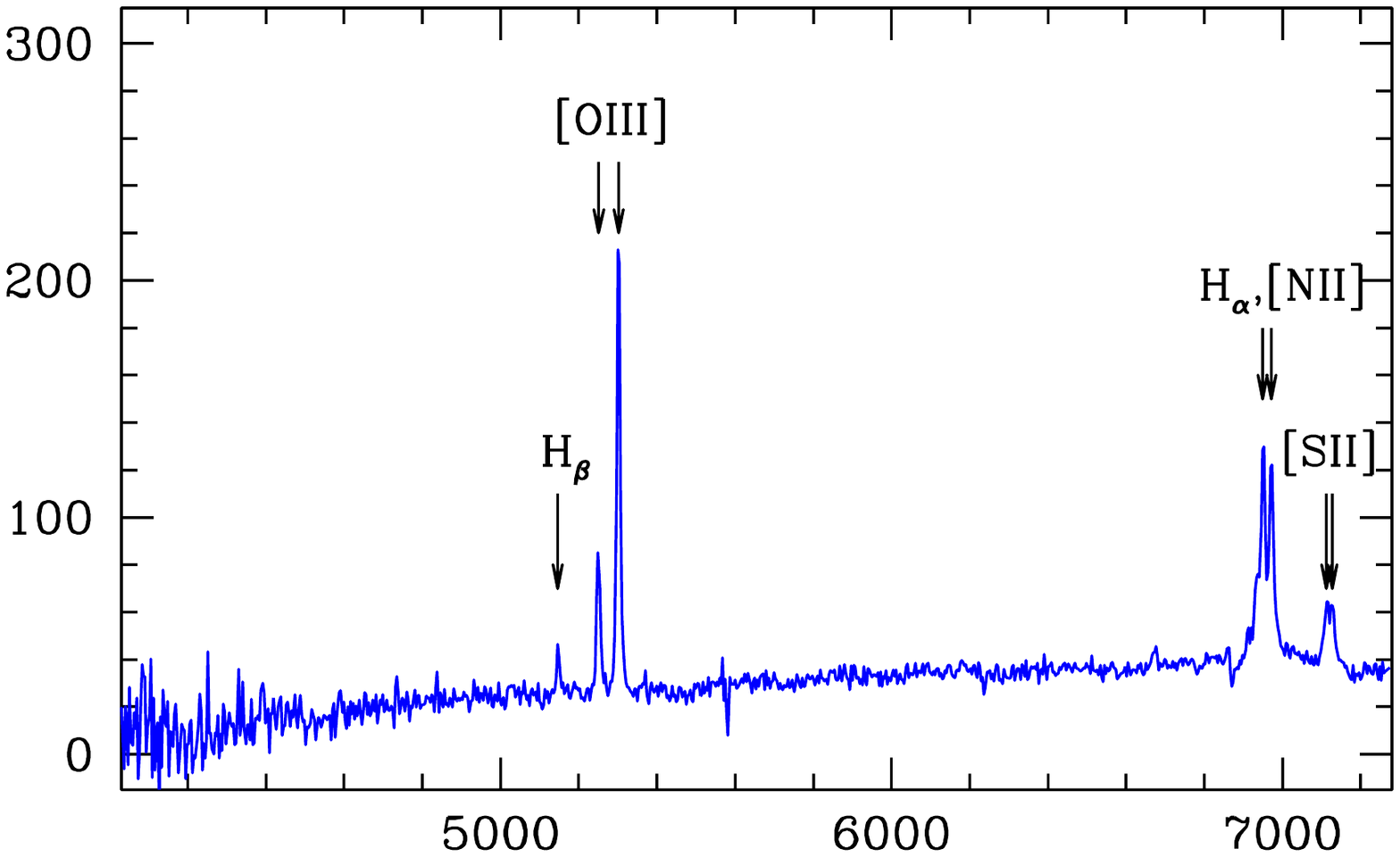}{$\lambda$, \AA}{Flux, $10^{-17}$ erg\,s$^{-1}$\,cm$^{-2}$}
  \end{floatrow}
  \bigskip

  \caption{Same as Fig.~\ref{fig:spec0432}, but for SRGA\,J$200431.6+610211$.}
  \label{fig:spec2004}
\end{figure*}

The X-ray source is reliably identified (Fig.~\ref{fig:spec2004}) with the galaxy 2MASX\,J$20043237\!+\!6102311$ = WISEA\,J$200432.40\!+\!610230.8$, whose infrared color ($W1-W2=0.89$) points to the possible presence of an active nucleus.

\begin{table*}
  \caption{Spectral features of SRGA\,J$200431.6\!+\!610211$ = 2MASX\,J$20043237\!+\!6102311$} 
  \label{tab:j2004}
  \vskip 2mm
  \renewcommand{\arraystretch}{1.1}
  \renewcommand{\tabcolsep}{0.35cm}
  \centering
  \footnotesize
  \begin{tabular}{lcccc}
    \noalign{\doubleline}
    Line & Wavelength, \AA& Flux, $10^{-16}$~erg\,s$^{-1}$\,cm$^{-2}$& Eq. Width, \AA& $FWHM$, km\,s$^{-1}$\\
    \noalign{\vskip 3pt\hrule\vskip 5pt}
    H${\beta}$ & 5147 & $18.0\pm 1.2$ & $-6.7\pm 0.5$ & $(4.7\pm 0.7)\times 10^2$\\
    O\,III$\lambda$4959 & 5250 & $57\pm 5$ & $-19\pm 4$ & $(4.8\pm 0.7)\times 10^2$\\
    O\,III$\lambda$5007 & 5302 & $196\pm 7$ & $-59\pm 3$ & $(4.8\pm 0.7)\times 10^2$\\
    H${\alpha}$ & 6950 & $110\pm 11$ & $-23\pm 2$ & $(4.7\pm 0.5)\times 10^2$\\
    N\,II$\lambda$6584 & 6972 & $111\pm 11$ & $-23\pm 2$ & $(5.0\pm 0.5)\times 10^2$\\
    SII$\lambda$6718 & 7113 & $31\pm 3$ & $-7.7\pm 0.6$ & $(5.5\pm 0.4)\times 10^2$\\
    SII$\lambda$6732 & 7129 & $31\pm 3$ & $-7.7\pm 0.6$ & $(5.2\pm 0.4)\times 10^2$\\
  \noalign{\vskip 3pt\hrule\vskip 5pt}
  \end{tabular}
  \begin{flushleft}
  \end{flushleft}
\end{table*}

Our optical observations were carried out on October 22, 2020, at the AZT-33IK telescope using VPHG600G. Five spectral images with an exposure time of 300 s each were obtained; the total exposure time was 25~min.

Narrow H${\beta}$, [O\,III]${\lambda}$4959, [O\,III]${\lambda}$5007, H${\alpha}$, [N\,II]${\lambda}$6584, and sulfur doublet emission lines are seen in our spectrum (Fig.~\ref{fig:spec2004}). The line characteristics are given in Table~\ref{tab:j2004}. The redshift was determined from seven lines: $z = 0.05866 \pm 0.00013$.

The ratios log([[O\,III]$\lambda 5007/$H${\beta})=1.04\pm 0.03$ and log([N\,II]$\lambda 6584/$H${\alpha})=0.00\pm 0.06$. From the position on the BPT diagram (Fig.~\ref{chart:bpt}) and the absence of broad lines, the object can be classified as a Seyfert 2 galaxy.

A slight absorption is detected in the object’s X-ray spectrum (Fig.~\ref{fig:spec2004}): $\nh\sim 5\times 10^{21}$~cm$^{-2}$.

\subsection{\bf SRGA\,J224125.9\!+\!760343}

This X-ray source was discovered during the \rosat\ all-sky survey: 2RXS\,J224124.5\!+\!760346 \citep{2rxs}, but its nature so far has remained unknown. The source was detected by both \art\ and \ero\ telescopes of the \srg\ observatory.

\begin{figure*}
  \centering
  \vfill
  SRGA\,J$224125.9\!+\!760343$
  \vfill
  \vskip 0.5cm
  \begin{floatrow}
    \includegraphics[width=0.40\columnwidth]{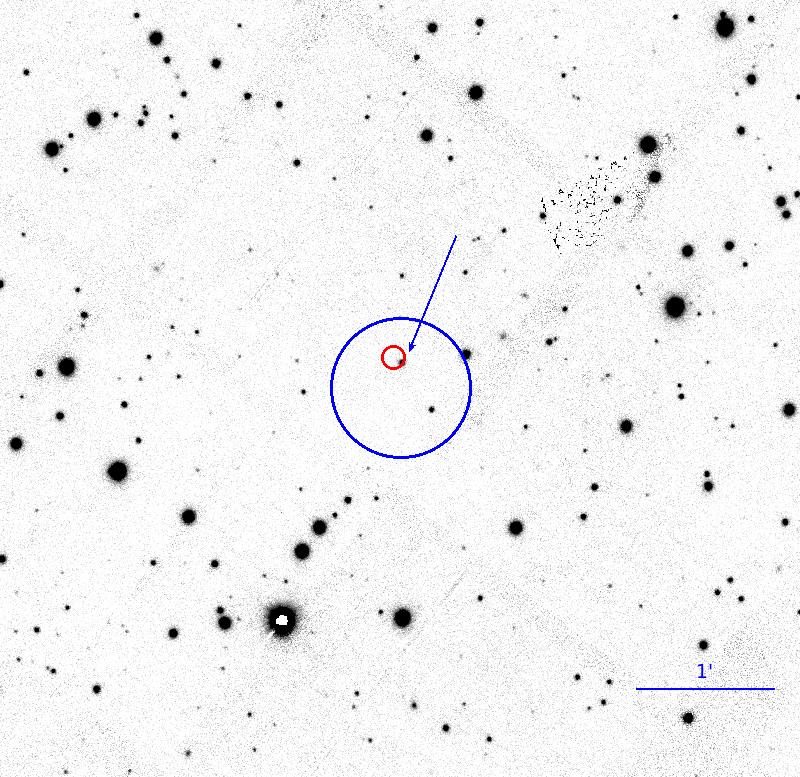}
    \includegraphics[width=0.45\columnwidth]{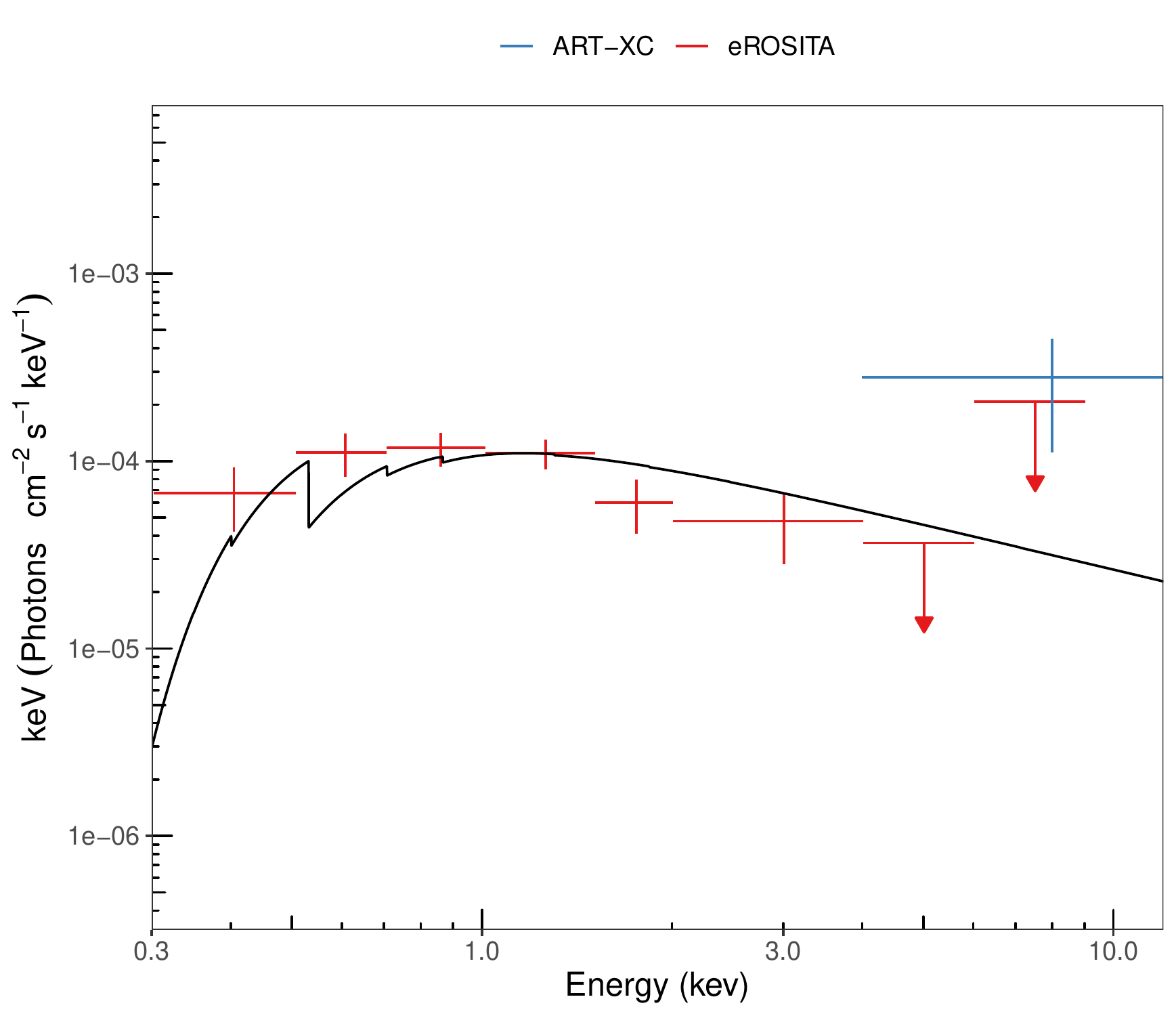}
  \end{floatrow}
  \vfill
  RTT-150
  \vfill
  \begin{floatrow}
    \smfigure{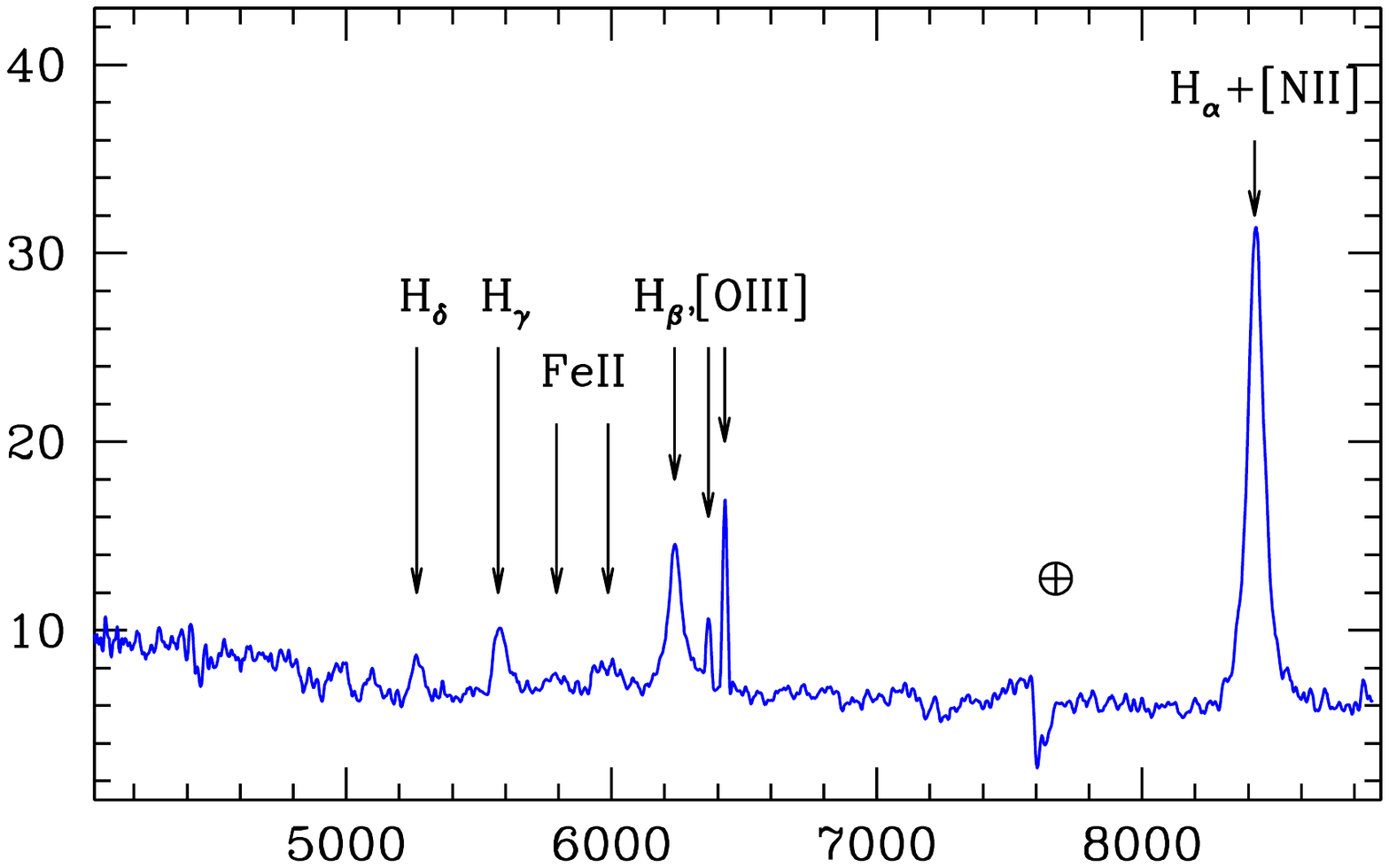}{$\lambda$, \AA}{Flux, $10^{-17}$ erg\,s$^{-1}$\,cm$^{-2}$}
  \end{floatrow}
  \bigskip

  \caption{Same as Fig.~\ref{fig:spec0432}, but for SRGA\,J$224125.9\!+\!760343$. In contrast to other sources, we show the power-law model with a slope $\Gamma=2.4$ that provides a better quality of the fit than does the model with the standard slope $\Gamma=1.8$.}
  \label{fig:spec2241}
\end{figure*}

The X-ray source is reliably identified (Fig.~\ref{fig:spec2241}) with the infrared source WISEA\,J$224125.79\!+\!760353.8$, whose infrared color ($W1-W2=0.97$) points to the possible presence of an active nucleus.

\begin{table*}
  \caption{Spectral features of SRGA\,J$224125.9\!+\!760343$ = WISEA\,J$224125.79\!+\!760353.8$} 
  \label{tab:j2241}
  \vskip 2mm
  \renewcommand{\arraystretch}{1.1}
  \renewcommand{\tabcolsep}{0.35cm}
  \centering
  \footnotesize
  \begin{tabular}{lcccc}
    \noalign{\doubleline}
    Line & Wavelength, \AA& Flux, $10^{-16}$\,erg\,s$^{-1}$ cm$^{-2}$& Eq. Width & $FWHM$, km\,s$^{-1}$\\
    \noalign{\vskip 3pt\hrule\vskip 5pt}
    H${\gamma}$, narrow & 5579 & $<0.8$ & $>-1.2$ & --\\
    H${\gamma}$, broad & 5579 & $17.4\pm 0.8$ & $-26\pm 1$ & $(2.1\pm 0.2)\times 10^3$\\
    FeII$\lambda$4570 & 5982 & $18.6\pm 1.3$ & -- & --\\
    H${\beta}$, narrow & 6239 & $4.4\pm 0.3$ & $-6.4\pm 0.4$ & $(3.8\pm 0.9)\times 10^2$\\
    H${\beta}$, broad & 6239 & $51\pm 1$ & $-35\pm 2$ & $(1.5\pm 0.2)\times 10^3$\\
    O\,III$\lambda$4959 & 6365 & $7.7\pm 0.2$ & $-11.3\pm 0.3$ & $(3.7\pm 0.9)\times 10^2$\\
    O\,III$\lambda$5007 & 6428 & $20.1\pm 0.3$ & $-29\pm 1$ & $(3.7\pm 0.9)\times 10^2$\\
    H${\alpha}$, narrow & 8429 & $6.0\pm 0.9$ & $-8.8\pm 1.3$ & $(2.8\pm 0.7)\times 10^2$\\
    H${\alpha}$, broad & 8429 & $197\pm 3$ & $-288\pm 5$ & $(2.3\pm 0.1)\times 10^3$\\
  \noalign{\vskip 3pt\hrule\vskip 5pt}
  \end{tabular}
\end{table*}

Our optical observations were carried out on June 21, 2020, at RTT-150. Three spectral images with an exposure time of 1800~s each were obtained; the total exposure time was 90~min.

The Balmer H${\alpha}$, H${\beta}$, H${\gamma}$, and H${\delta}$, emission lines with narrow and broad components are seen in our spectrum (Fig.~\ref{fig:spec2241}). The H${\alpha}$ line merged with the [N\,II]$\lambda$6548 and [N\,II]$\lambda$6584 lines. Obviously, for this reason, the measured $FWHM$ of the broad H${\alpha}$ component slightly exceeds the $FWHM$ of the corresponding H${\beta}$ component. The [O\,III]$\lambda$4960, [O\,III]$\lambda$5007 emission lines and the complex of Fe\,II$\lambda$4570 ($\lambda 4434$--$\lambda 4684$) lines are also present in the spectrum. The characteristics of all lines are presented in Table~\ref{tab:j2241}. The redshift determined from six emission lines is $z=0.2834 \pm 0.0003$. The narrow-line flux ratio is log([O\,III]$\lambda 5007/$H${\beta})=0.66$, while the ratio log([N\,II]$\lambda 6584/$H${\alpha})$ is difficult to estimate due to the line merging. The relative narrowness of the broad Balmer line components ($FWHM$ (H${\beta})<2000$~km\,s$^{-1}$) and the presence of a noticeable Fe\,II emission suggest that this object is a narrow-line Seyfert 1 galaxy.

There is no evidence of an additional absorption in the object’s X-ray spectrum (Fig.~\ref{fig:spec2241}), except for the absorption in our Galaxy. At a fixed slope of the power-law spectrum $\Gamma=1.8$ we obtain a strict upper limit on the internal absorption: $N_{\rm H} < 4\times 10^{20}$~cm$^{-2}$.

\subsection{\bf SRGA\,J232446.8\!+\!440756}

This X-ray source was discovered in the 4--12 keV band by the \art\ telescope of the \srg\ observatory and, at the same time, was not detected in softer X-rays by the \ero\ telescope.

\begin{figure*}
  \centering
  \vfill
  SRGA\,J$232446.8\!+\!440756$
  \vfill
  \vskip 0.5cm
  \begin{floatrow}
    \includegraphics[width=0.40\columnwidth]{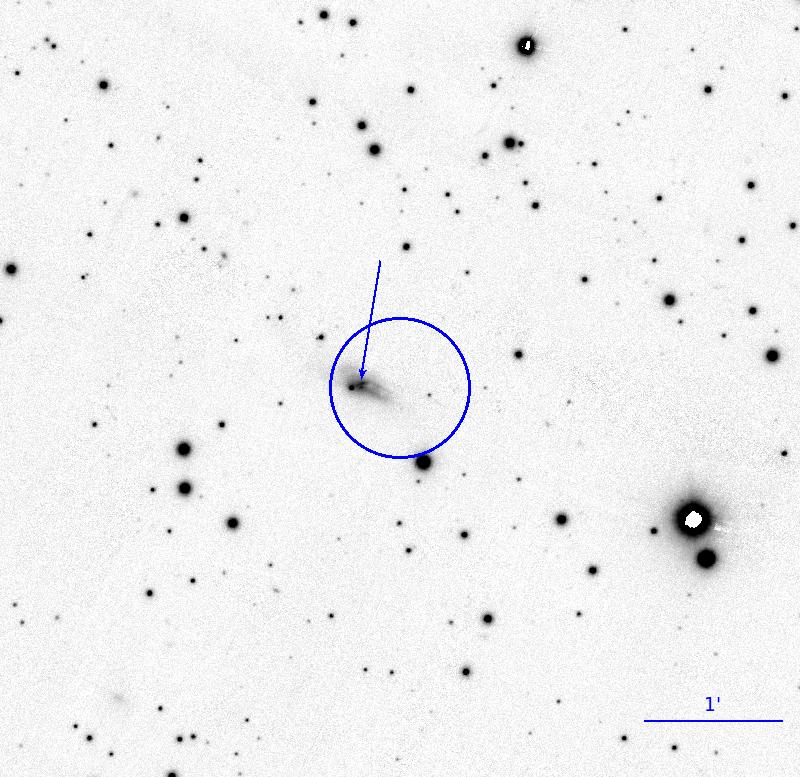}
    \includegraphics[width=0.40\columnwidth]{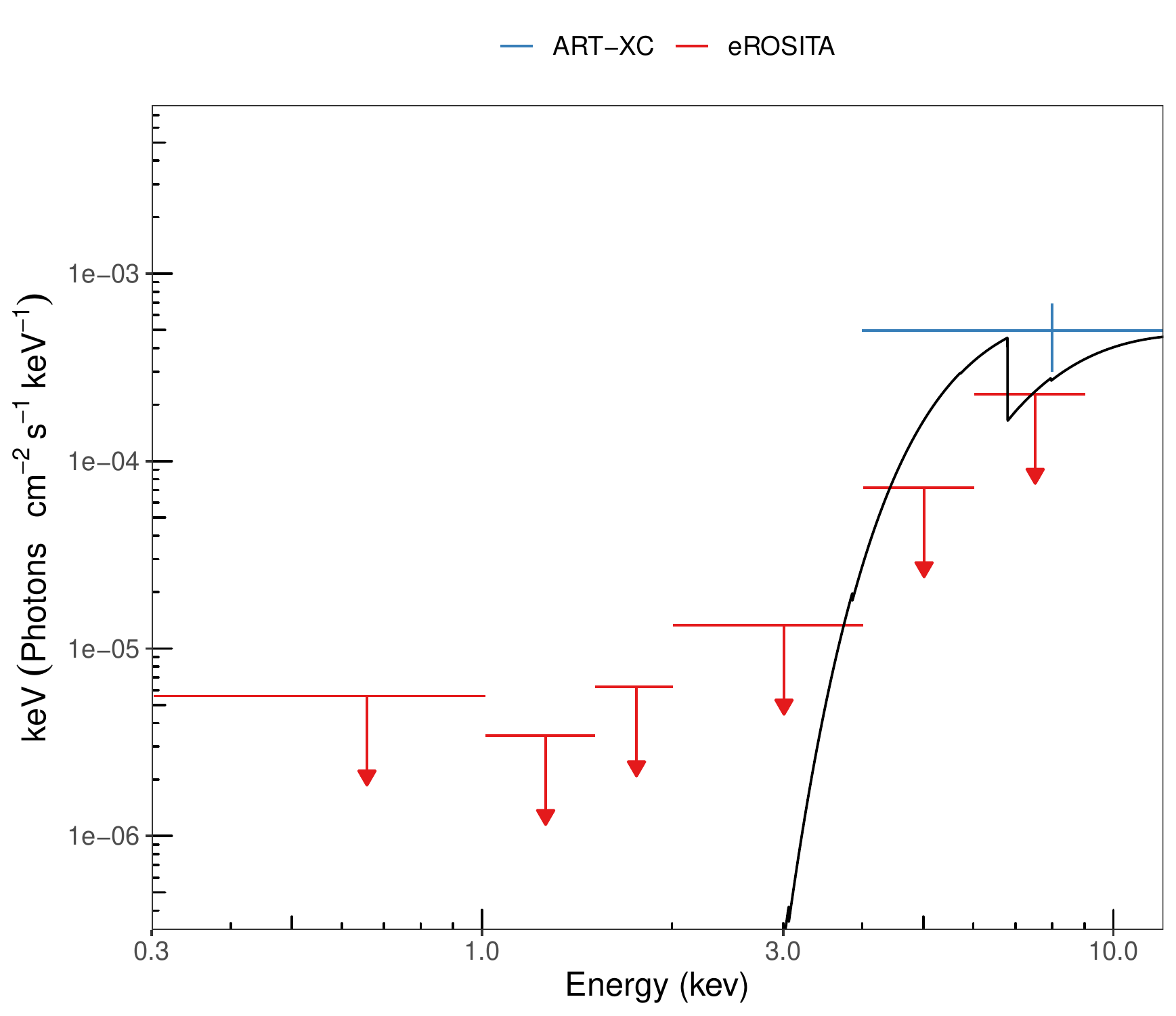}
  \end{floatrow}
  \vfill
  RTT-150
  \vfill
   \begin{floatrow}
       \smfigure{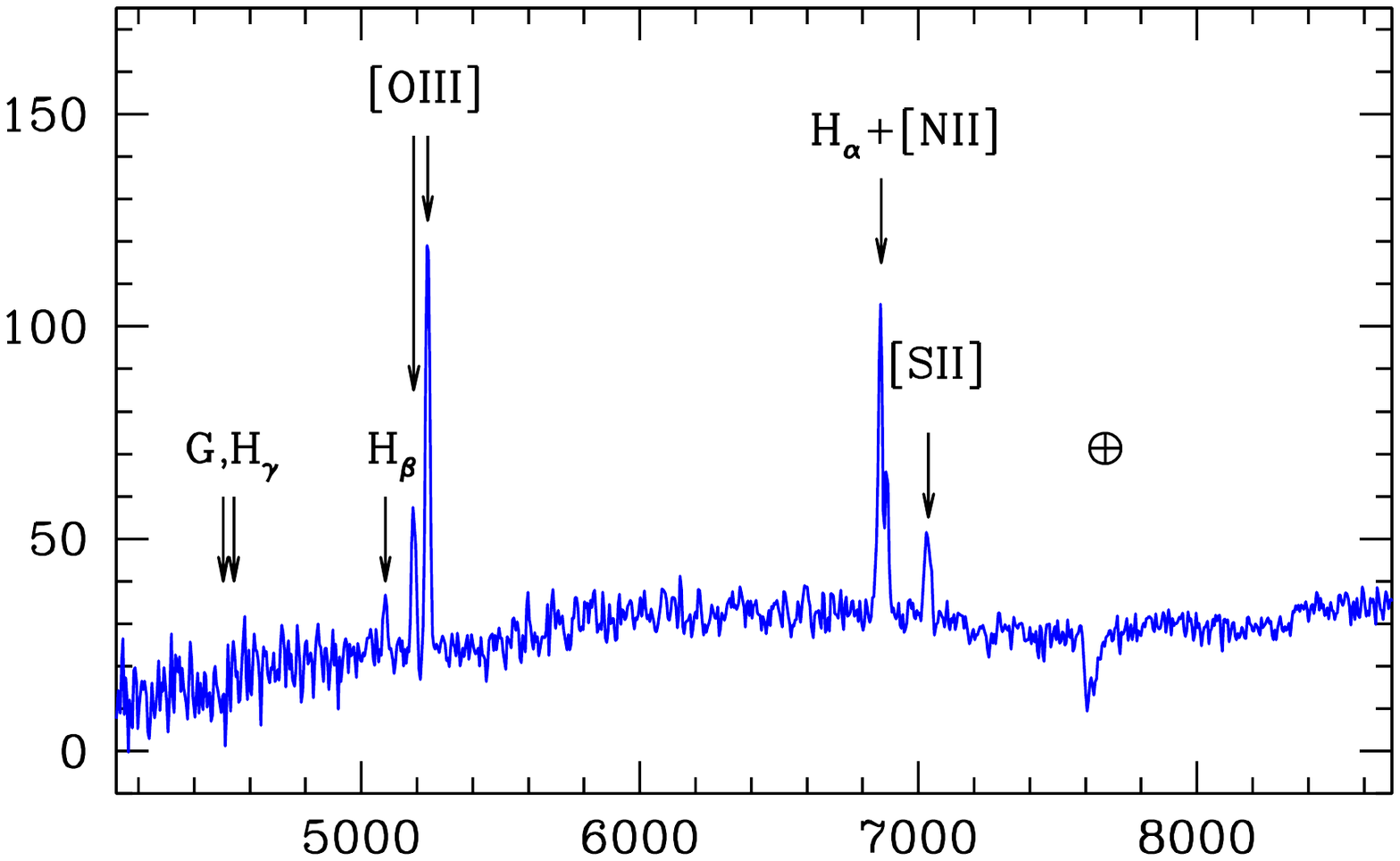}{$\lambda$, \AA}{Flux, $10^{-17}$ erg\,s$^{-1}$\,cm$^{-2}$}
     \end{floatrow}
  \bigskip

  \caption{Same as Fig.~\ref{fig:spec0057}, but for SRGA\,J$232446.8\!+\!440756$.}
  \label{fig:spec2324}
\end{figure*}

The X-ray source can be identified with the irregular galaxy 2MASX\,J$23244834\!+\!4407564$ = WISEA\,J$232448.36\!+\!440756.5$ (Fig.~\ref{fig:spec2324}). Its redshift is known: $z=0.04634$ \citep{2mass}, while its infrared color ($W1-W2=0.83$) points to the presence of an active nucleus. However, the galaxy has not yet been classified as an AGN from optical spectroscopy.

\begin{table*}
  \caption{Spectral features of SRGA\,J$232446.8\!+\!440756$ = 2MASX\,J$23244834\!+\!4407564$} 
  \label{tab:j2324}
  \vskip 2mm
  \renewcommand{\arraystretch}{1.1}
  \renewcommand{\tabcolsep}{0.35cm}
  \centering
  \footnotesize
  \begin{tabular}{lcccc}
    \noalign{\doubleline}
    Line & Wavelength, \AA& Flux, $10^{-16}$ erg\,s$^{-1}$\,cm$^{-2}$& Eq. Width, \AA& $FWHM$, km\,s$^{-1}$\\
    \noalign{\vskip 3pt\hrule\vskip 5pt}
    H${\beta}$ & 5087 & $27\pm 4$ & $-12\pm 3$ & $(7.4\pm 1.0)\times 10^2$\\
    O\,III$\lambda$4959 & 5189 & $74\pm 9$ & $-42\pm 9$ & $(6.8\pm 1.0)\times 10^2$\\
    O\,III$\lambda$5007 & 5239 & $196\pm 12$ & $-102\pm 17$ & $(7.3\pm 1.0)\times 10^2$\\
    H${\alpha}$ & 6865 & $134\pm 9$ & $-41\pm 6$ & $(5.3\pm 0.8)\times 10^2$\\
    N\,II$\lambda$6584 & 6887 & $46\pm 7$ & $-17\pm 3$ & $(1.5^{+3.7}_{-1.5})\times 10^2$\\
  \noalign{\vskip 3pt\hrule\vskip 5pt}
  \end{tabular}
\end{table*}

Our optical observations were carried out on June 10, 2020, at RTT-150. Nine spectral images with an exposure of 600~s each were obtained; the total exposure time was 90~min.

Narrow H${\alpha}$, H${\beta}$, [O\,III]$\lambda 4959,\lambda 5007$, [N\,II]$\lambda 6584$ emission lines and the [S II] doublet are seen in our spectrum (Fig.~\ref{fig:spec2324}). The line characteristics are given in Table~\ref{tab:j2324}. The redshift was determined from five emission lines and is $z = 0.04624 \pm 0.00020$, consistent with the previously measured value by \cite{2mass}. The ratios log([N\,II]$\lambda 6584/$H${\alpha})=-0.46\pm 0.07$ and log([O\,III]$\lambda 5007/$H${\beta})= 0.86\pm 0.07$. From the position on the BPT diagram (Fig.~\ref{chart:bpt}) and the absence of broad lines, the object can be classified as a Seyfert 2 galaxy.

The non-detection by the \ero\ telescope in combination with the 4--12 keV flux measured by the \art\ telescope (Fig.~\ref{fig:spec2324}) allows a strict upper limit to be placed on the absorption column density: $\nh>3\times 10^{23}$~cm$^{-2}$.

\section{PROPERTIES OF THE DETECTED AGNs}

Table \ref{tab:list2} presents the main properties of the AGNs that we managed to identify in this study. Apart from the redshift and the optical type, the estimated column density of cold matter inside the object NH and its X-ray luminosity $L_{\rm X}$ in the 4--12 keV energy band are given for each object.

We estimated the X-ray luminosity based on the flux in the 4--12 keV energy band measured by the \art\ telescope of the \srg\ observatory and the photometric distance to the object calculated from its redshift. The presented values of $L_{\rm X}$ neglect the k-corrections and were not corrected for the line of sight absorption. The first of these corrections should not be significant, given the low redshifts of the objects and that the slope of the AGN X-ray spectra does not differ greatly from $\Gamma = 2$. As regards the absorption correction, although it may turn out to be large for three objects from the sample with a high column density ($\nh > 10^{23}$~cm$^{-2}$), it is virtually impossible to reliably take into account based on the existing \art\ and \ero\ data (there are too few detected photons). Therefore, it should be kept in mind that the true luminosity of these heavily obscured AGNs can be greater than that given in the table by several times.

As can be seen from Table \ref{tab:list2}, most of the objects being discussed are Seyfert galaxies with a luminosity $\lx\sim 10^{42}$--$10^{44}$~erg\,s$^{-1}$ in the nearby Universe ($z < 0.1$), except for the source SRGA\,J$224125.9\!+\!760343$ at $z = 0.28$ with a luminosity $\lx\sim 10^{45}$~erg\,s$^{-1}$ that, using the traditional terminology, may be attributed to quasars.

Almost all of the investigated objects fall into the region of Seyfert galaxies on the standard BPT diagram (Fig. \ref{chart:bpt}) of the [O\,III]$\lambda$5007/H$\beta$ and [N\,II]$\lambda$6584/H$\alpha$ flux ratios. SRGA\!J$005751.0\!+\!210846$ was not placed on this diagram, because the required information about the emission lines cannot be obtained from the available optical spectra. In this case, we are dealing with a galaxy (LEDA 1643776) seen edge-on, so that the line emission regions in its active nucleus can be completely hidden from the observer. SRGA\!J$224125.9\!+\!760343$ did not fall on the BPT diagram either, because the broad H$\alpha$ component merged with the [N\,II]$\lambda$6584 line and, for this reason, it is impossible to estimate the line parameters. Undoubtedly, both objects are AGNs, because they are characterized by a high X-ray luminosity. As has already been discussed above, the object SRGA\!J$014157.0\!-\!032915$ is located on the BPT diagram in the region of star-forming galaxies, but near the region of Seyfert galaxies. The narrow emission lines in its spectrum probably result not only from the accretion of matter onto the SMBH in the galactic nucleus, but also from intense star formation in the galaxy.

Six of the eight investigated object(if SRGA\,J$005751.0\!+\!210846$ seen edge-on is included) turned out to be Seyfert 2 or intermediate type (1.9) galaxies. The detection of an appreciable absorption in their X-ray spectra is quite expectable.

One of the objects (SRGA\,J$224125.9\!+\!760343$) turned out to be a narrow-line Seyfert 2 galaxy. We can estimate the SMBH mass in this object from the formula \citep{vandp06}

\begin{eqnarray*}
\lg{\mbh}&=& \lg\left[\left(\frac{FWHM({\rm H}\beta)}{1000~{\rm km/s}}\right)^2\left(\frac{L({\rm H}\beta)}{10^{42}~{\rm erg/s}}\right)^{0.63}\right]. \\
    && +6.67.
\end{eqnarray*}

In our case, $FWHM({\rm H}\beta)=(1.5\pm 0.2)\times 10^3$~km\,s$^{-1}$ and the line flux is $F({\rm H}\beta)=(1.30\pm 0.02)\times 10^{-14}$~erg\,s$^{-1}$~cm$^{-2}$ (see Table.~\ref{tab:j2241}), which allows the line luminosity to be estimated at $z = 0.2834$, $L({\rm H}\beta)\approx 3.3\times 10^{42}$~erg\,s$^{-1}$. As a result, we find $\mbh\approx 2.3\times 10^7M_\odot$.

For such a relatively small black hole the critical Eddington luminosity is $\ledd\approx 3\times 10^{45}$~erg\,s$^{-1}$. At the same time, the measured luminosity of the source SRGA\!J$224125.9\!+\!760343$ in the X-ray energy band (4--12~keV) is $\lx\sim (2-13)\cdot 10^{44}$\,erg\,s$^{-1}$. Since the bolometric luminosity $\lbol$ of an AGN usually exceeds the X-ray luminosity at least by several times \citep[see, e.g.,][]{sazonov04}, we conclude that $\lbol\sim \ledd$ for SRGA\,J$224125.9\!+\!760343$. This corresponds to the universally accepted paradigm \citep[see, e.g.,][]{mathur} that in narrow-line Seyfert 1 galaxies the accretion of matter occurs at a rate close to the critical one.

\begin{table*}
  \caption{Properties of the AGNs} 
  \label{tab:list2}
  \vskip 2mm
  \renewcommand{\arraystretch}{1.1}
  \renewcommand{\tabcolsep}{0.35cm}
  \centering
  \footnotesize
  \begin{tabular}{ccccc}
    \noalign{\doubleline}     
    Object & Optical type$^1$ & $z$ & $\nh ^2$ & $\log\lx ^3$ \\
  \hline
  SRGA\,J$005751.0\!+\!210846$ & Sy2$^4$ & $0.04798\pm 0.00002$ & $>1\times10^{3}$ & $43.7_{-0.3}^{+0.2}$ \\ 
  SRGA\,J$014157.0\!-\!032915$ & Sy2 & $0.01878\pm 0.00003$ & $>3\times10^{2}$ & $42.5_{-1.2}^{+0.3}$ \\ 
  SRGA\,J$043209.6\!+\!354917$ & Sy1 & $0.0506\pm 0.0010$ & $ 3.0_{-0.7}^{+ 0.8}$ &  $43.8_{-0.3}^{+0.2}$\\ 
  SRGA\,J$045049.8\!+\!301449$ & Sy1.9 & $0.03308\pm 0.00004$ & $38_{-10}^{+11}$&$43.4_{-0.3}^{+0.2}$  \\ 
  SRGA\,J$152102.3\!+\!320418$ & Sy2 & $0.1143\pm 0.0003$ & $25_{-6}^{+ 6}$ & $44.1_{-0.4}^{+0.2}$ \\ 
  SRGA\,J$200431.6\!+\!610211$ & Sy2 & $0.05866\pm 0.00013$ & $ 4.7_{-1.4}^{+ 2.2}$ & $43.6_{-0.3}^{+0.2}$ \\ 
  SRGA\,J$224125.9\!+\!760343$ & NLSy1 & $0.2834\pm 0.0004$ & $<0.4$ & $44.9_{-0.6}^{+0.2}$ \\ 
  SRGA\,J$232446.8\!+\!440756$ & Sy2 & $0.0462\pm 0.0002$ & $>3\times10^2$ & $43.5_{-0.3}^{+0.2}$ \\ 
   \hline

    \noalign{\vskip 3pt\hrule\vskip 5pt}
  \end{tabular}
  \begin{flushleft}
    $^1$ Sy1, Sy1.9, and Sy2 are Seyfert 1, 1.9, and 2 galaxies, respectively, NLSy1 is a narrow-line Seyfert 1 galaxy.
    
    $^2$ In units of $10^{21}$\,cm$^{-2}$, the errors and limits correspond to the 90\% confidence level, while the 68\% confidence level is presented for the source SRGA\,J$014157.0\!-\!032915$.
    
    $^3$ The luminosity uncorrected for absorption in the observed 4--12 keV energy band in units of erg\,s$^{-1}$.
    
    $^4$ The classification is arbitrary, because the galaxy is seen edge-on.

  \end{flushleft}
 \end{table*}

\begin{figure*}
  \centering
  \vfill
  \vskip 2cm
  \begin{minipage}{0.90\columnwidth}
    \centering
    \includegraphics[width=0.78\columnwidth]{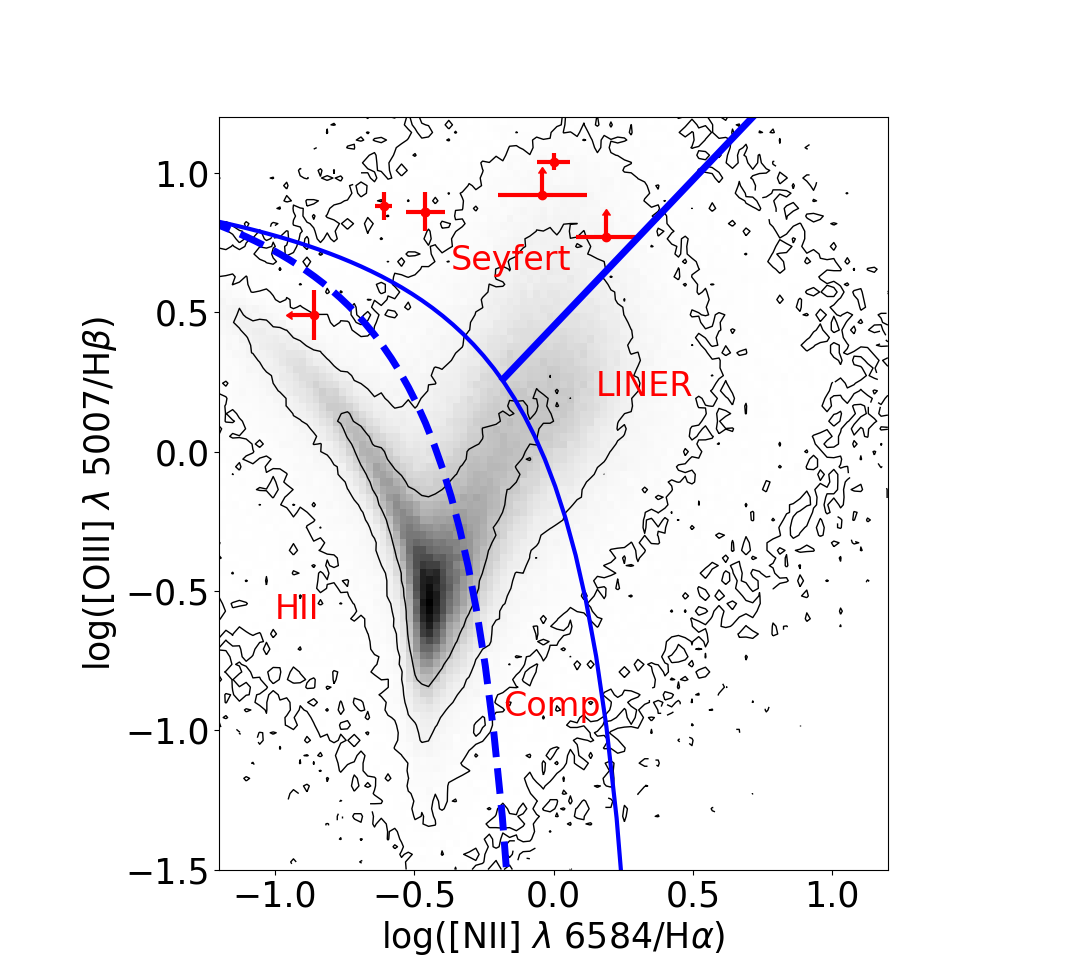}
  \end{minipage} 
  \label{chart:bpt}
  \bigskip

  \caption{Positions of the AGNs under study (red dots and limits) on the BPT diagram \citep{bpt} constructed from SDSS data (release 7, \citealt{sdssdr7}). The demarcation lines between different classes of galaxies were taken from \cite{kauff03} -- the dashed line, \cite{kewly01} -- the thin line, and \cite{scha07} -- the thick line. Only the six objects for which we managed to determine the parameters of the required lines are shown. The diagram was constructed with the help of the site http://wwwmpa.mpa-garching.mpg.de/SDSS/DR7/Data/gal\_line\_dr7\_v5\_2.fit.gz.}
  \label{chart:bpt}
\end{figure*}

\section{CONCLUSIONS}

We have identified eight new AGNs among the X-ray sources detected during the first all-sky survey with the \art\ telescope aboard the \srg\ observatory. We measured the redshifts of these objects and studied their optical and X-ray properties. Most of the objects turned out to be Seyfert 2 galaxies and exhibit significant absorption in the X-ray spectrum. For three AGNs, the absorption column density exceeds $3 \times 10^{23}$\,cm$^{-2}$. For this reason, they are detected only in fairly hard X-rays with the \art\ telescope and are not detected in softer X-rays with \ero. In one of these objects (SRGA\,J$005751.0\!+\!210846$) the absorption may be associated mainly with the interstellar gas in the host galaxy seen edge-on. One of the objects (SRGA\,J$224125.9\!+\!760343$) turned out to be a narrow-line Seyfert 1 galaxy with a luminosity close to the Eddington limit.

The results of this study confirm the expectations that the \art\ telescope is an efficient instrument for searches of heavily obscured and other interesting AGNs in the relatively nearby ($z \lesssim 0.3$) Universe. The \srg\ all-sky survey will last for more than three years more, which will allow a lot of such objects to be discovered.

\section{ACKNOWLEDGMENTS}

This work was supported by RSF grant no. 19-12-00396. We thank T\"{U}BITAK, the Space Research Institute of the Russian Academy of Sciences, the Kazan Federal University, and the Academy of Sciences of Tatarstan for supporting the observations at the Russian–Turkish 1.5-m telescope (RTT-150). The measurements with the AZT-33IK telescope were performed within the basic financing of the FNI II.16 program and were obtained using the equipment of the Angara sharing center\footnote{http://ckp-rf.ru/ckp/3056/}. The work of I.F. Bikmaev, E.N. Irtuganov, and E.A. Nikolaeva was supported by the subsidy (project no. 0671-2020-0052) allocated to the Kazan Federal University for a State assignment in the sphere of scientific activities. In this study we used observational data from the \ero\ telescope onboard the \srg\ observatory. The \srg\ observatory was built by Roskosmos in the interests of the Russian Academy of Sciences represented by its Space Research Institute (IKI) within the framework of the Russian Federal Space Program, with the participation of the Deutsches Zentrum f\"{u}r Luft- und Raumfahrt (DLR). The \srg/\ero\ X-ray telescope was built by a consortium of German Institutes led by MPE, and supported by DLR. The \srg\ spacecraft was designed, built, launched, and is operated by the Lavochkin Association and its subcontractors. The science data are downlinked via the Deep Space Network Antennae in Bear Lakes, Ussurijsk, and Baykonur, funded by Roskosmos. The \ero\ data used in this work were processed using the \emph{eSASS} software system developed by the German \ero\ consortium and the proprietary data reduction and analysis software developed by the Russian \ero\ Consortium.

\end{document}